\documentclass{aa}
\usepackage{txfonts}
\usepackage{graphicx}
\usepackage{natbib}
\usepackage{longtable}
\bibpunct{(}{)}{;}{a}{}{,}
\begin{document}
\title{Statistical analyses of long--term variability of AGN at high radio frequencies}

\author{T. Hovatta \inst{1} \and M. Tornikoski \inst{1} \and M. Lainela \inst{2}\and H.J. Lehto \inst{2,3} \and E. Valtaoja \inst{2,3} \and I. Torniainen \inst{1} \and M.F. Aller \inst{4} \and H.D. Aller \inst{4}}
\institute{Mets\"ahovi Radio Observatory, Helsinki University of Technology, Mets\"ahovintie 114, 02540 Kylm\"al\"a, Finland  \\ \email{tho@kurp.hut.fi} \and Tuorla Observatory, University of Turku, V\"ais\"al\"antie 20, 21500 Piikki\"o, Finland \and Department of Physics, University of Turku, 20100 Turku, Finland \and Department of Astronomy, University of Michigan, Ann Arbor, MI 48109, USA}
\date{Received 23 March 2007 / Accepted 23 April 2007}
\abstract
{}
{We present a study of variability time scales in a large sample of 
Active Galactic Nuclei at several frequencies between 4.8 and 230\,GHz. We 
investigate the differences of various AGN types and frequencies and 
correlate the measured time scales with physical parameters such as 
the luminosity and the Lorentz factor. 
Our sample consists of both high and low polarization quasars, 
BL Lacertae objects and radio galaxies. 
The basis of this work is the 22\,GHz, 37\,GHz and\,87 GHz monitoring data from
the Mets\"ahovi Radio Observatory spanning over 25 years. In
addition, we used higher 90\,GHz and 230\,GHz frequency data obtained with the SEST-telescope between 1987 and 2003. 
Further lower frequency data at 4.8\,GHz, 8\,GHz and 14.5\,GHz from the University of Michigan monitoring programme have been used.
}
{We have applied three different statistical methods to study the 
time scales: The structure function, the discrete correlation function 
and the Lomb--Scargle periodogram. We discuss also the differences and relative merits of these three methods.}
{Our study reveals that smaller flux density variations occur in these 
sources on 
short time scales of 1-2 years, but larger outbursts happen quite rarely, 
on the average only once in every 6 years. 
We do not find any significant 
differences in the time scales
between the source classes. The time scales are also only weakly 
related to the luminosity suggesting that the shock formation 
is caused by jet instabilities rather than the central black hole.
}
{}

\keywords{Galaxies: active -- Methods: statistical}
\maketitle
\section{Introduction}

Long term multifrequency monitoring data of a large sample of 
Active Galactic Nuclei (AGNs)
provides an efficient means for studying the physical processes behind the variability behaviour of individual objects. 
It is also a useful tool for studying 
differences between various AGN classes.
\cite{hughes92} used the structure function (SF) to study the time scales 
of variability in
a large sample of sources at frequencies 4.8, 8 and 14.5\,GHz, 
using data from the University of Michigan monitoring
programme. The SF was also used by \cite{lainela93}, hereafter Paper I,
to study the time
scales of the Mets\"ahovi monitoring sample. Since then the amount of
data has more than tripled. In this paper we analyse the updated
extensive database and compare the results 
to Paper I.

The discrete correlation function (DCF) and
the Lomb--Scargle periodogram (LS--periodogram) were also used to search for the 
variability time scales at several frequency bands. Both of these have 
been used 
previously to study periodicities and time scales in individual sources
\citep[eg.][]{villata04, ciaramella04, raiteri03, raiteri01,
roy00}. \cite{aller03} studied the Pearson-Readhead extragalactic source
sample, monitored at the University of Michigan, by
using the LS--periodogram without finding any significant
periodicities.

We have 
used these methods to look for the 
typical flare occurrence rates and other variability 
properties
of this large sample of sources. By using more than one
method we 
hoped
to ensure that the time scales obtained 
are real.
Furthermore, the experience we gain in using different methods will let us
choose in the future the proper methods for specific analysis needs.
The DCF was used to study the radio--optical correlations in the 
Mets\"ahovi monitoring sample by \cite{tornikoski94b} and \cite{hanski02}.
In the present paper
we have, however, used the autocorrelation at each
radio frequency band instead of cross correlation between 
different frequencies.

The long term variability time scales can tell us about how often certain 
objects, or certain classes of objects, are in a flaring state and 
how long do these flares typically last. 
We also learn about
flare evolution from one frequency domain to the 
other. This 
helps us in developing radio shock models.
The knowledge of typical flare time scales
is also important for 
the work done on extragalactic foreground sources for the
ESA Planck Surveyor 
satellite\footnote{http://www.rssd.esa.int/Planck} mission, to be
launched in 2008. The Planck satellite 
will be used to study the cosmic 
microwave background (CMB) emission, and all the foreground sources, 
including AGNs, 
must be removed from the results.  Therefore it is important to understand 
the characteristic time scales of 
AGN variability at high radio frequencies. This paper is part of our 
broader study of various AGN types that affect the foreground of Planck.

This paper is organised as follows: In Sect. \ref{sec:sample}
we describe the source sample and the data. 
The methods used for the analyses are described in Sect. \ref{sec:methods} and
the results are presented in Sect. \ref{sec:results}. Finally,
we will discuss the scientific outcome of the results in 
Sect. \ref{sec:discussion}. In Sect. \ref{sec:conclusions} we will 
draw the conclusions.
 
\section{The Sample and Observations}\label{sec:sample}
The sample consists of 80 sources which have been selected from the
Mets\"ahovi monitoring sample. We included objects for which we have data
from a time window of over 10 years in at least two frequency bands.
However, most of the sources have been monitored 
up to 25 years,
enabling a search for longer time scales. Sources which are included in the 
monitoring list are bright sources with 
a flux density of
least 1 Jy in the active state. 
The present sample is shown in Table \ref{table:sourcelist}\footnote{Table \ref{table:sourcelist} is available only in the 
electronical edition of the journal, www.aanda.org}. The columns 
show the observing frequency, the length of the time series 
and the number of points for each source.

The sample consists of different types of AGNs. 24 of the sources are
BL Lacertae objects (BLOs), 23 Highly Polarized Quasars (HPQs),
28 Low Polarization Quasars (LPQs) and 5 Radio Galaxies (GALs).
Quasars are considered to be highly polarized if
their optical polarization has exceeded 
3 percent at some point in the past. It is
possible that some of the low polarized objects are in reality HPQs 
but they have not been observed in an active state. For 5 
objects we had no information about their optical polarization, and they 
are considered LPQs in this study. When we study the statistical 
differences of the various groups, only BLOs, HPQs and LPQs will be considered
, because of the low number of GALs in our sample.
The variability of BLOs will also be studied in more detail in a forthcoming 
paper by Nieppola et al. (2007, in preparation for A\&A).

The core of this work is the monitoring data from the 14 meter Mets\"ahovi Radio 
Telescope. We have been monitoring a sample of AGNs for over 25 years 
at frequencies 22, 37 and 87\,GHz
\citep{salonen87, terasranta92, terasranta98, terasranta04, terasranta05}.
Our study also includes unpublished data at 37 GHz from December 2001 
to April 2005. The data for BL Lacertae objects from this period 
are published in \cite{nieppola07}.
The observation method and data reduction process are 
described in \cite{terasranta98}.
The Swedish--ESO Submillimetre Telescope (SEST) at La Silla, Chile, was 
used in our monitoring campaign to sample the high frequency, 90 and 230\,GHz 
variability of southern and equatorial sources
(\cite{tornikoski96}, Tornikoski et al. 2007, in preparation for A\&A).
The monitoring campaign at 
SEST lasted from 1987 to 2003. 
High frequency data at 90 and 230\,GHz were also collected from the
literature \citep{Steppe88, Steppe92, Steppe93, Reuter97}.
The lower frequency data at frequencies 4.8, 8 and 14.5\,GHz were provided
by the University of Michigan Radio Observatory (UMRAO) monitoring programme. Details of calibration and data reduction are described in \cite{aller85}.
We had sufficiently well--sampled 230\,GHz data for our analysis 
for only 7 sources and therefore that
frequency band is not used when average time scales and differences between 
the source classes are studied.

The median interval between the observations in individual sources varied 
from 31 days to 47 days at 22 and 90\,GHz, respectively. 
At 37\,GHz the median sampling rate was 41 days, but the 
value depends on the source.
The minimum average value, 6.8 days, was at 37\,GHz for the source 3C 84,
which is used as a secondary calibrator in the Mets\"ahovi observations. 
The minimum average for a source not used as a calibrator was 8.9 days for 
the source 3C273 at 37\,GHz. The maximum average value, 186.4 days, was for 
the source 2234+282 also at 37\,GHz. 
We also compared the sampling rates in Paper I for the 40 sources in common in 
our samples 
with sampling rates from data after 1993. At 22\,GHz the difference is larger
with a median of 39 days in Paper I and 19 days after 1993. At 37\,GHz the 
median sampling rate is 31 days for both data sets. 
For the 40 sources not included 
in Paper I, the median sampling rate during the whole period was 
40 days at 22\,GHz and 61 days at 37\,GHz.

\section{Methods}\label{sec:methods}
Three methods were used to study the characteristic
time scales of different types of AGNs: SF, DCF and 
LS--periodogram. We chose to use three different methods because
we also wanted to study these methods in more detail, as well as the 
differences between them. An additional reason
for using the SF analysis was to compare the 
results with those of the analysis done in Paper I. This way we can study how 
13 years of additional data affect the time scales.

\subsection{The structure function}\label{sec:SF}
 The general description of the structure function is given by 
\cite{simonetti85}. We will use only the first-order SF defined in 
Eq. \ref{eq:sf},

\begin{equation} \label{eq:sf}
 D^1(\tau) = \left<\left[S(t) - S(t + \tau)\right]^2\right>
\end{equation}
where
$S(t)$ is the flux density at time $t$ and $\tau$ is the time lag. Our analysis follows 
the descriptions in Paper I and \cite{hughes92}. Here we will only 
shortly describe the method.

\begin{figure}
\resizebox{\hsize}{!}{\includegraphics{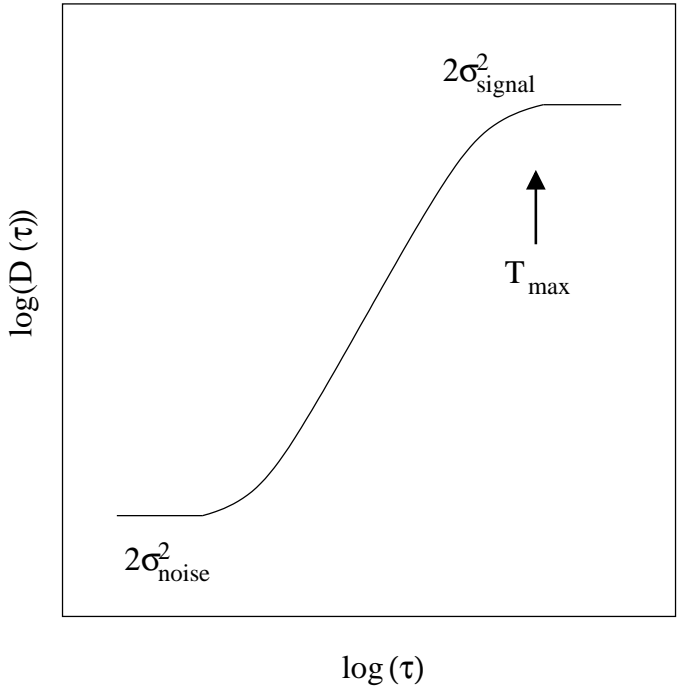}}
\caption{Ideal Structure Function}
\label{ideal_sf}
\end{figure}

An ideal structure function is presented in Fig. \ref{ideal_sf}. 
It consists of
two plateaus and a slope between them. The $x$-axis shows logarithm of the timelag, $\tau$, and the $y$-axis shows the logarithm of the structure function, $D(\tau)$. We can identify a time scale 
at the point $T_\mathrm{max}$ where the structure function reaches its 
second, higher plateau. This time scale is the maximum time scale of correlated
behaviour. For lags longer than $T_\mathrm{max}$, we have a plateau with 
amplitude equal to twice the variance of the signal. The lower plateau at short timelags is equal to twice the average variance of the measurement noise for a single data point.

In addition, we can find out the nature of the process 
from the slope b between the two plateaus. 
If the lightcurve can be modelled as a white or a red noise process, then
a slope of unity in the structure function implies shot noise, and a slope 
close to zero implies flicker or white noise. Usually the process is a mixture 
of these processes and the slope is something between 0 and 1. See 
\cite{hufnagel92} for details. 
If one large outburst dominates the time series, the slope may be steeper 
than 1. A strong linear trend or a strong periodic oscillation in 
the data is expected to produce a slope of 2.

In our analysis the timelag runs from 1 week to the length of the light curve. Many of the sources have been observed approximately once a week, 
and therefore
we have chosen the lower limit to be one week. 
Higher frequencies  (90 and 230\,GHz) have usually been monitored for shorter times.
We first calculated the differences squared for all the two point pairs,
and then to create the structure function we averaged all the samples into 0.1 dex wide bins.

We estimated the error caused by observational uncertainties to the SF by an
independent bootstrap method. For each source and frequency we created 
a model light curve by running over the light curve a boxcar with a length 
of 10 days and averaging. We tested also other averaging lengths but using 
much shorter values we would have not provided enough non-zero residuals 
for scrambling. Much longer averaging lengths would have started to average 
over significant variability in some sources. We then subtracted this 
10-day model from each light curve and created a bank of residuals. 
A new simulated light curve was made by adding to each point in the model 
a randomly selected residual. The time sampling of the original light curve 
was thus preserved. This was repeated so that each residual was selected 
once. Using this new light curve we recalculated a new simulated SF. 
The procedure was repeated 1000 times. This enables us to put confidence 
limits to the SF, e.g. the 99\% confidence limit at a given time scale was 
at the value where only 5 points were above or below the value. This 
method clearly does not require the confidence limits to be symmetric.

Results of the analysis are presented in Sect. \ref{sec:results} 
and comparison with Paper I in 
Sect. \ref{sec:discussion}.

\subsection{Discrete correlation function}\label{sec:dcf}
Discrete correlation function was first introduced by \cite{edelson88}.
\cite{hufnagel92} generalized the method to include a better error estimate.
The advantage of DCF compared to other correlation methods is that it is 
suitable for unevenly sampled data, which is usually the case in 
astronomical observations. 
Here we will
describe only briefly the method and formulae used, and refer to 
\cite{tornikoski94b} and \cite{hufnagel92} for details.

First we need to calculate the unbinned correlations for the time 
series. Note that the formulation below allows for cross correlations.
This is done using Eq. \ref{eq:udcf}, 
where $a_i$ and $b_j$ are individual points in the time series $a$ and $b$, 
$\bar{a}$ and $\bar{b}$ are the means of the time series, and
$\sigma^{2}_{a}$ and $\sigma^{2}_{b}$ are the variances. After 
calculating the UDCF the 
correlation function is binned. 
The method does not define {\sl a priori}
the bin size so we have tested several values. If the bin size is 
too large, information is lost. On the other hand, if the bin size is too 
small, we can get spurious correlations, and the time scales may be 
difficult to 
interpret.
We have chosen a bin size of 50 days for all 
autocorrelations. For several sources we also tested smaller bin size of 
25 days but this did not make noticeable changes to the results.

\begin{equation} \label{eq:udcf}
UDCF_\mathrm{ij} = \frac{(a_i-\bar{a})(b_j-\bar{b})}{\sqrt{\sigma^{2}_{a}\sigma^{2}_{b}}}
\end{equation}

By binning the UDCF we obtain the DCF using Eq. \ref{eq:dcf}.
Here $\tau$ is the time 
of the centre of the time bin
and $n$ is the number of 
points in each bin. We can also calculate the error in each bin by using
Eq. \ref{eq:dcf_error}. This represents the standard deviation of the UDCF estimates within the bin. 

\begin{equation} \label{eq:dcf}
DCF(\tau) = \frac{1}{n}\Sigma UDCF_\mathrm{ij}(\tau)
\end{equation}

\begin{equation} \label{eq:dcf_error}
\sigma_\mathrm{dcf}(\tau) = \frac{1}{n-1}\left\{\Sigma\left[UDCF_\mathrm{ij}-DCF\left(\tau\right)\right]^2\right\}^{0.5}
\end{equation}

A disadvantage of this method is that it does not give any exact probability 
value for the calculated results. The only way we can study the reliability 
of the method is to use simulations. 
The error caused by observational uncertainties have been estimated 
with the same bootstrap method as for SF, described in the previous section.
Here we have used 10000 simulations which means that the 99\% 
confidence limit was at the value where 50 points were above or below the 
value.
One should note that these confidence limits represent the ambiguity 
caused by observational uncertainties and do not address possible 
questions posed by the sampling of the data. The errors obtained with this 
method are similar to those calculated using Eq. \ref{eq:dcf_error}.

We also used simulated periodic data to test the 
capability of DCF to find real time 
scales and found out that it could detect all real time scales well.
For our simulations we created flux density curves with strict periodicities by 
multiplying flares of real sources to extend over a period of 25 years. 
The DCF could detect the period for the simulated data with good precision.
Results of the DCF analysis are presented in Sect. \ref{sec:results}.

\subsection{Lomb-Scargle periodogram}\label{sec:ls}
Fourier--based methods can be used for studying periodicities in light 
curves. We tested if these methods are also suitable for studying the 
characteristic variability time scales of AGNs. We have chosen 
the commonly 
used method of Lomb-Scargle periodogram for this study \citep{lomb76,
scargle82}. It is based on the discrete Fourier-transform
which has been modified for unevenly sampled data. The method searches
for sinusoidal periodicities in the frequency domain. This turns out
to be problematic because our light curves are not well represented by 
a sum of sinusoidal functions. Usually the most significant spike of the 
periodogram turned out to be at the time scale of the total length of 
the time series, and other spikes 
were its harmonics.

We have taken the formulae as they appear in \cite{press92}. First 
we need to calculate the mean and the standard
deviation of the time series. 
We calculated the periodogram with a sampling interval of $1/4T$ in the 
frequency space. Here $T$ is the total length of the time series. The 
upper limit to which the periodogram was calculated was $N/2T$, where $N$ 
is the total number of observations. In evenly spaced data this would 
correspond to the Nyquist frequency. 
Now we can calculate the Lomb-Scargle periodogram 
by using Eq. \ref{eq:period},

\begin{eqnarray}\label{eq:period}
P_N(\omega) & = & \frac{1}{2\sigma^{2}} \frac{\left[
\Sigma_{j}(a_{j}-\bar{a})\cos \omega
(t_{j}-\tau)\right] }{\Sigma_{j} \cos^{2} \omega
(t_{j}-\tau)}\\
\nonumber & & +\frac{1}{2\sigma^{2}} \frac{\left[
\Sigma_{j}(a_{j}-\bar{a})\sin \omega
(t_{j}-\tau)\right] }{\Sigma_{j} \sin^{2} \omega
(t_{j}-\tau)}
\end{eqnarray}

where $t_j$ is the date of an individual observation and $\omega$ is the 
frequency at which we are calculating the periodogram. $a_j$ is an 
individual data point of time series $a$, and $\bar{a}$ is the mean of the 
time series. $\tau$ can be calculated
from Eq. \ref{eq:tau}.
\begin{equation} \label{eq:tau}
tan(2\omega\tau) = \frac{\Sigma_j\sin2\omega t_j}{\Sigma_j\cos2\omega t_j}
\end{equation}
A false alarm probability level of $z\approx\ln(N/p)$ for a 99.9\%
for the most significant spike can be calculated \citep{scargle82}.
Unfortunately, this does not tell anything about the 
significance of other spikes in the periodogram and therefore only 
the most significant one is used in this analysis. 
In radio data the annual gaps in the data are much shorter than in the optical
and do not contribute to aliasing in a significant 
way creating spurious spikes.

\section{Results}\label{sec:results}
\subsection{Results of the structure function analysis}
We used the structure function to study the characteristic time scales of 
80 sources at frequencies 4.8, 8, 14.5, 22, 37, 90 and 230\,GHz.

In total we calculated 411 structure functions from which we could
determine 447 time scales. 
In 39 cases we could not determine a variability time scale because 
the function was too flat or because the errors in the structure function 
were too large.
In 60 cases we could only get a lower limit for the time scale.
We could also determine more than one time scale in 69 cases.

The relative number of sources for which we could not determine a time 
scale depended on the frequency, for example at 14.5\,GHz we had only 
two such cases (sources 1147+245 and 0446+112), but at 90\,GHz 
one third of the sources failed to provide us with a good 
estimate for the time scale.
This was  mainly due to the undersampling at 90\,GHz and especially 
at 230\,GHz, which caused
the structure functions to have large errors and therefore to be 
difficult to interpret. Also there were only five sources with data 
at 230\,GHz.

Also the number of the lower limit estimates varied between the 
frequency bands and the source classes. In Table \ref{table:SF_averages}
the number of such sources is written in parenthesis for each frequency band
and class.
The relative percentage of lower limit time scales in LPQs, BLOs and HPQs were
24\%, 13\% and 11\%. 

\begin{figure*}
\includegraphics[width=17cm]{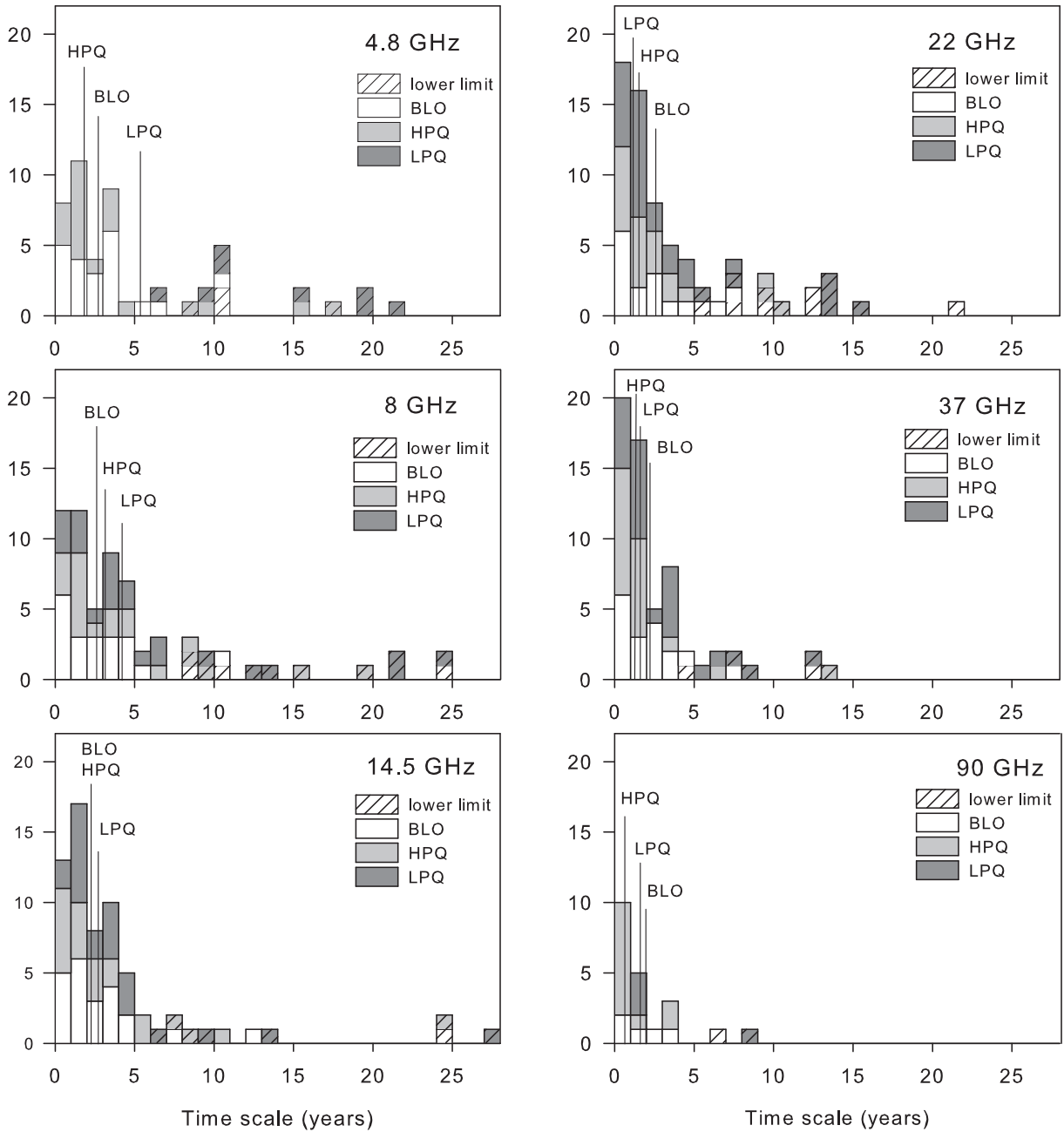}
\caption{Distributions of time scales from structure function analysis at all frequency bands and source classes. The median values for each source class are shown by vertical lines. The time scales are in the observer's frame.}
\label{histo_SF}
\end{figure*}

We have plotted the distributions of the time scales at different 
frequencies and source classes in Fig. \ref{histo_SF}. The 
median time scales of each class in the histograms are marked 
by vertical lines.
The average and median time scales are presented in 
Table \ref{table:SF_averages}.

\begin{table*}
\caption[]{Averages and median values of the Structure Function analysis. Number of sources, for which only a lower limit time scale could be determined, is written in parenthesis.}
\label{table:SF_averages}
\centering
\begin{tabular}{rrrrrrrrrr}
\hline
\hline
Freq 	  & 	 type 	  & 	 ALL 	  & 	 number  	  & 	 BLO 	  & 	 number  	  & 	 HPQ 	  & 	 number  	  & 	 LPQ 	  & 	 number  	  \\
  	  & 	  	  & 	[years]  	  & 	 of sources 	  & [years]	  	  & 	 of sources 	  & 	[years]  	  & 	 of sources 	  & 	[years]  	  & 	 of sources 	  \\
\hline
\hline
 4.8 	  & 	 average 	  & 	 5.2 	  & 	 66(12) 	  & 	 3.6 	  & 	 23(2) 	  & 	 4.2 	  & 	 19(2) 	  & 	 7.8 	  & 	 20(8) 	  \\
  	  & 	 redshift corr. 	  & 	 2.6 	  & 	 65 	  & 	 2.8 	  & 	 22 	  & 	 2.1 	  & 	 19 	  & 	 3.8 	  & 	 20 	  \\
  	  & 	 median 	  & 	 3.0 	  & 	 66 	  & 	 2.7 	  & 	 23 	  & 	 1.9 	  & 	 19 	  & 	 5.3 	  & 	 20 	  \\
  	  & 	 redshift corr. 	  & 	 1.7 	  & 	 65 	  & 	 1.8 	  & 	 22 	  & 	 1.0 	  & 	 19 	  & 	 1.9 	  & 	 20 	  \\
\hline
 8 	  & 	 average 	  & 	 5.6 	  & 	 70(13) 	  & 	 4.2 	  & 	 23(3) 	  & 	 4.8 	  & 	 20(4) 	  & 	 7.0 	  & 	 22(6) 	  \\
  	  & 	 redshift corr. 	  & 	 3.5 	  & 	 69 	  & 	 3.4 	  & 	 22 	  & 	 2.4 	  & 	 20 	  & 	 3.6 	  & 	 21 	  \\
  	  & 	 median 	  & 	 3.2 	  & 	 70 	  & 	 2.7 	  & 	 23 	  & 	 3.1 	  & 	 20 	  & 	 4.1 	  & 	 22 	  \\
  	  & 	 redshift corr. 	  & 	 1.8 	  & 	 69 	  & 	 1.7 	  & 	 22 	  & 	 1.6 	  & 	 20 	  & 	 1.8 	  & 	 21 	  \\
\hline
 14.5 	  & 	 average 	  & 	 4.3 	  & 	 70(8) 	  & 	 3.7 	  & 	 23(1) 	  & 	 4.0 	  & 	 21(3) 	  & 	 4.5 	  & 	 22(4) 	  \\
  	  & 	 redshift corr. 	  & 	 2.7 	  & 	 69 	  & 	 3.0 	  & 	 22 	  & 	 2.0 	  & 	 21 	  & 	 2.2 	  & 	 22 	  \\
  	  & 	 median 	  & 	 2.3 	  & 	 70 	  & 	 2.2 	  & 	 23 	  & 	 2.2 	  & 	 21 	  & 	 2.7 	  & 	 22 	  \\
  	  & 	 redshift corr. 	  & 	 1.3 	  & 	 69 	  & 	 1.3 	  & 	 22 	  & 	 0.8 	  & 	 21 	  & 	 1.2 	  & 	 22 	  \\
\hline
 22 	  & 	 average 	  & 	 4.1 	  & 	 74(14) 	  & 	 5.0 	  & 	 21(6) 	  & 	 2.9 	  & 	 20(2) 	  & 	 2.1 	  & 	 28(6) 	  \\
  	  & 	 redshift corr. 	  & 	 2.5 	  & 	 73 	  & 	 4.1 	  & 	 20 	  & 	 1.3 	  & 	 20 	  & 	 1.9 	  & 	 28 	  \\
  	  & 	 median 	  & 	 1.9 	  & 	 74 	  & 	 2.7 	  & 	 21 	  & 	 1.5 	  & 	 20 	  & 	 1.1 	  & 	 28 	  \\
  	  & 	 redshift corr. 	  & 	 1.1 	  & 	 73 	  & 	 2.4 	  & 	 20 	  & 	 0.7 	  & 	 20 	  & 	 1.1 	  & 	 28 	  \\
\hline
 37 	  & 	 average 	  & 	 2.6 	  & 	 66(6) 	  & 	 2.8 	  & 	 19(2) 	  & 	 2.0 	  & 	 19(1) 	  & 	 3.9 	  & 	 23(3) 	  \\
  	  & 	 redshift corr. 	  & 	 1.5 	  & 	 65 	  & 	 2.1 	  & 	 18 	  & 	 1.1 	  & 	 19 	  & 	 1.4 	  & 	 23 	  \\
  	  & 	 median 	  & 	 1.4 	  & 	 66 	  & 	 2.2 	  & 	 19 	  & 	 1.2 	  & 	 19 	  & 	 1.5 	  & 	 23 	  \\
  	  & 	 redshift corr. 	  & 	 0.7 	  & 	 65 	  & 	 1.3 	  & 	 18 	  & 	 0.5 	  & 	 19 	  & 	 0.8 	  & 	 23 	  \\
\hline
 90 	  & 	 average 	  & 	 2.0 	  & 	 23(2) 	  & 	 2.5 	  & 	 6(1) 	  & 	 1.1 	  & 	 11(0) 	  & 	 3.2 	  & 	 4(1) 	  \\
  	  & 	 redshift corr. 	  & 	 1.3 	  & 	 23 	  & 	 1.8 	  & 	 6 	  & 	 0.6 	  & 	 11 	  & 	 1.9 	  & 	 4 	  \\
  	  & 	 median 	  & 	 1.1 	  & 	 23 	  & 	 2.0 	  & 	 6 	  & 	 0.7 	  & 	 11 	  & 	 1.6 	  & 	 4 	  \\
  	  & 	 redshift corr. 	  & 	 0.6 	  & 	 23 	  & 	 1.2 	  & 	 6 	  & 	 0.4 	  & 	 11 	  & 	 1.1 	  & 	 4 	  \\
\hline
\end{tabular}
\end{table*}

Because a substantial number of time scale estimates are lower 
limits, a much
better representative for a characteristic time scale is the median time 
scale of the group. Practically all median time scales have value of less 
than 5 years.
The median time scales also shorten on average with increasing frequency. 

To provide a measure of the slope of the structure function between the 
plateaus we calculated the local slope in each structure function using 
trains of 2, 11 and 20 points.  By comparing the 
plots of slope vs.\ time scale we estimated the actual slope for each light 
curve. The 2 point slope provided a local measure of the slope, while 
for the overall slope between the plateaus, the 11 point or the 20 point 
trains gave more reliable results. Which of the two was better depends 
on the distance of the plateaus in $\log\tau$. This was possible in 358 cases. 
The slopes varied from 0 to 2.2 while the average of all the frequency bands
was near 1, which is 
expected if the light curve can be modelled as a $1/f^2$-type shot noise. 
At 37\,GHz the average slope 
was 0.72 and at 22\,GHz 0.83. We ran the 
Kruskal-Wallis analysis to 
see if there were differences between the slopes of the source classes. 
(All Kruskal-Wallis analyses in this paper have been performed with 
the Unistat software, version 5.0.)
At 4.8\,GHz we found that the BLOs differed from other classes significantly 
at P$>$95\% level. 
At 22\,GHz the HPQs differed from the BLOs significantly. 
At other frequencies we could not find any statistically significant 
differences between the groups. 

\subsection{Results of the DCF analysis}
\begin{figure*}
\includegraphics[width=17cm]{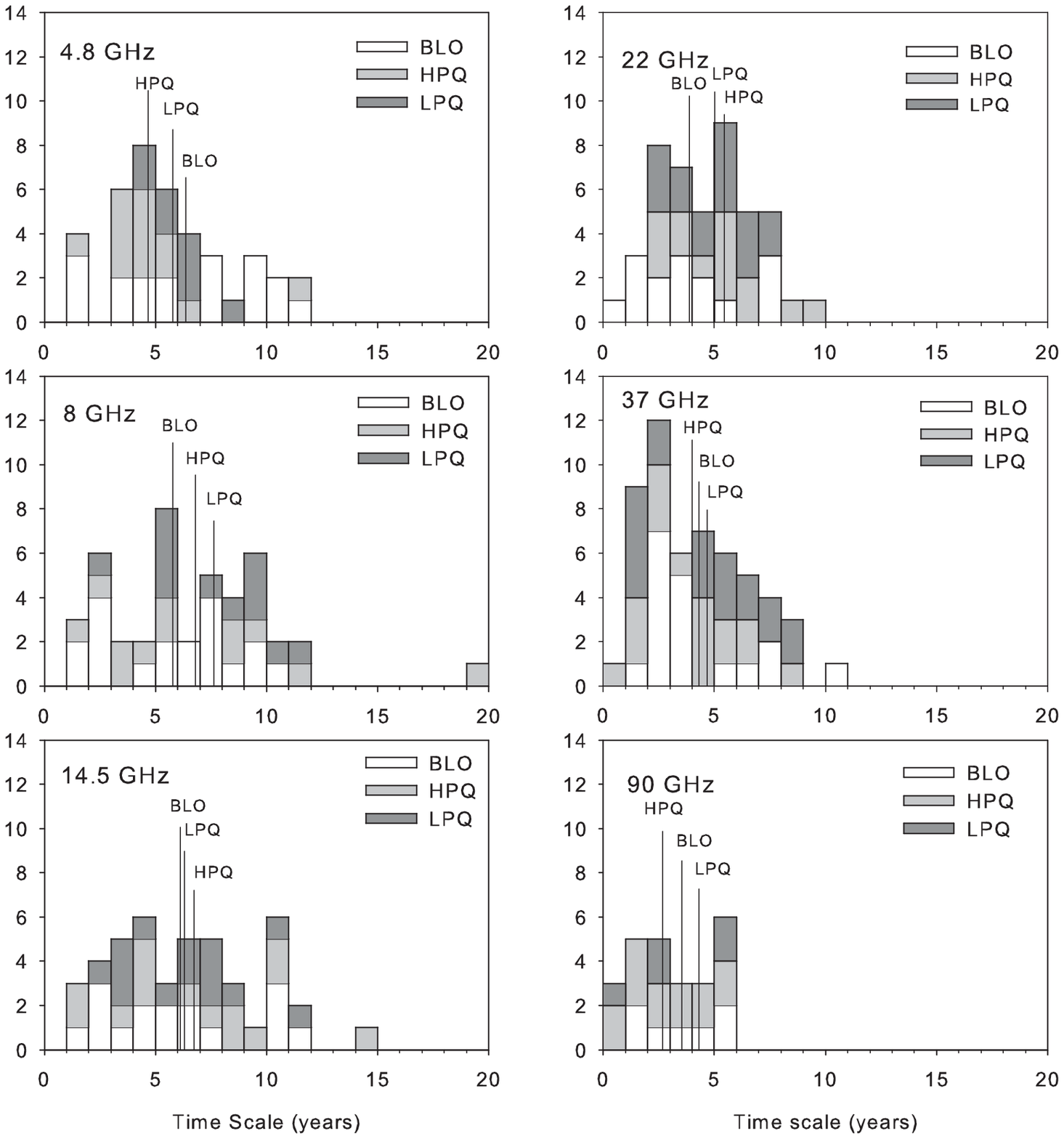}
\caption{Distributions of the most significant DCF time scales at all frequency bands. The averages of each source class are shown by vertical lines. The time scales are in the observer's frame.}
\label{histo_DCF}
\end{figure*}

We used the Discrete Correlation Function to calculate the
autocorrelation for 80 sources and obtained 411 autocorrelations.  
For many sources, more than one time scale was present in the DCF. 
For each we determined two time scales wherever possible. We identify as the 
most significant time scale the one that shows significant positive 
correlation after the DCF has been on the negative side. 
We were able to determine 273 such time scales. 
We also identified a time scale that we call 
the shortest time scale.  This is the first peak in the correlation function 
before it has gained negative values. It did not occur in all cases and usually
they appeared as small bumps in the DCF. We could determine the shortest 
time scale in 175 cases. Furthermore we calculated a redshift corrected 
time scale of the most significant time scale. 
Figure \ref{histo_DCF} shows the distribution of time scales at 
different frequency bands and source classes. 

\begin{table*}
\caption[]{Averages of DCF analysis time scales. For each frequency band the most significant, shortest and redshift corrected most significant time scale averages are shown. 
}
\label{table:DCF_averages}
\centering
\begin{tabular}{rrrrrrrrrr}
\hline
\hline
Freq 	  & 	 type 	  & 	 ALL 	  & 	 number 	  & 	 BLO 	  & 	 number 	  & 	 HPQ 	  & 	 number 	  & 	 LPQ 	  & 	 number 	  \\
 	  & 	  	  & 	[years]   & 	 of sources 	  &  [years]	  	  & 	 of sources 	  & [years]	  	  & 	 of sources 	  & [years]	  	  & 	 of sources 	  \\
\hline
\hline
 4.8 	  & 	 most signif. 	  & 	 5.9 	  & 	 42 	  & 	 6.3 	  & 	 18 	  & 	 4.7 	  & 	 13 	  & 	 5.9 	  & 	 8 	  \\
  	  & 	 shortest 	  & 	 2.3 	  & 	 35 	  & 	 2.3 	  & 	 13 	  & 	 2.1 	  & 	 9 	  & 	 2.3 	  & 	 11 	  \\
  	  & 	 redshift corr. 	  & 	 4.0 	  & 	 41 	  & 	 4.7 	  & 	 17 	  & 	 2.6 	  & 	 13 	  & 	 3.4 	  & 	 8 	  \\
\hline 
8 	  & 	 most signif. 	  & 	 6.7 	  & 	 45 	  & 	 5.9 	  & 	 19 	  & 	 6.9 	  & 	 12 	  & 	 7.6 	  & 	 12 	  \\
  	  & 	 shortest 	  & 	 2.0 	  & 	 25 	  & 	 1.8 	  & 	 10 	  & 	 1.9 	  & 	 7 	  & 	 2.4 	  & 	 8 	  \\
  	  & 	 redshift corr. 	  & 	 4.1 	  & 	 44 	  & 	 4.2 	  & 	 18 	  & 	 4.2 	  & 	 12 	  & 	 3.6 	  & 	 12 	  \\
\hline 
14.5 	  & 	 most signif. 	  & 	 6.5 	  & 	 48 	  & 	 6.1 	  & 	 16 	  & 	 6.8 	  & 	 14 	  & 	 6.2 	  & 	 14 	  \\
  	  & 	 shortest 	  & 	 1.9 	  & 	 31 	  & 	 2.3 	  & 	 11 	  & 	 1.9 	  & 	 11 	  & 	 1.9 	  & 	 8 	  \\
  	  & 	 redshift corr. 	  & 	 4.0 	  & 	 47 	  & 	 4.4 	  & 	 15 	  & 	 3.6 	  & 	 14 	  & 	 3.1 	  & 	 14 	  \\
\hline
22 	  & 	 most signif. 	  & 	 4.8 	  & 	 49 	  & 	 3.9 	  & 	 15 	  & 	 5.3 	  & 	 14 	  & 	 5.0 	  & 	 16 	  \\
  	  & 	 shortest 	  & 	 1.5 	  & 	 35 	  & 	 1.4 	  & 	 11 	  & 	 1.4 	  & 	 7 	  & 	 1.6 	  & 	 14 	  \\
  	  & 	 redshift corr. 	  & 	 2.9 	  & 	 48 	  & 	 2.9 	  & 	 14 	  & 	 2.7 	  & 	 14 	  & 	 2.7 	  & 	 14 	  \\
\hline 
37 	  & 	 most signif. 	  & 	 4.2 	  & 	 58 	  & 	 4.2 	  & 	 18 	  & 	 4.0 	  & 	 17 	  & 	 4.7 	  & 	 19 	  \\
  	  & 	 shortest 	  & 	 1.9 	  & 	 35 	  & 	 1.6 	  & 	 8 	  & 	 2.2 	  & 	 10 	  & 	 1.8 	  & 	 14 	  \\
  	  & 	 redshift corr. 	  & 	 2.6 	  & 	 57 	  & 	 3.2 	  & 	 17 	  & 	 2.0 	  & 	 17 	  & 	 2.4 	  & 	 19 	  \\
\hline
 90 	  & 	 most signif. 	  & 	 3.1 	  & 	 26 	  & 	 3.5 	  & 	 7 	  & 	 2.8 	  & 	 13 	  & 	 4.2 	  & 	 5 	  \\
  	  & 	 shortest 	  & 	 1.1 	  & 	 12 	  & 	 1.1 	  & 	 3 	  & 	 1.2 	  & 	 6 	  & 	 1.1 	  & 	 2 	  \\
  	  & 	 redshift corr. 	  & 	 1.9 	  & 	 26 	  & 	 2.3 	  & 	 7 	  & 	 1.6 	  & 	 13 	  & 	 1.6 	  & 	 5 	  \\
\hline 
\end{tabular}
\end{table*}

Table \ref{table:DCF_averages} shows
the average time scales from the DCF analysis
for the different frequency bands, and also for 
BLOs, HPQs and LPQs separately at each frequency band. 

\subsection{Results of the Lomb-Scargle periodogram analysis}
\begin{figure*}
\includegraphics[width=17cm]{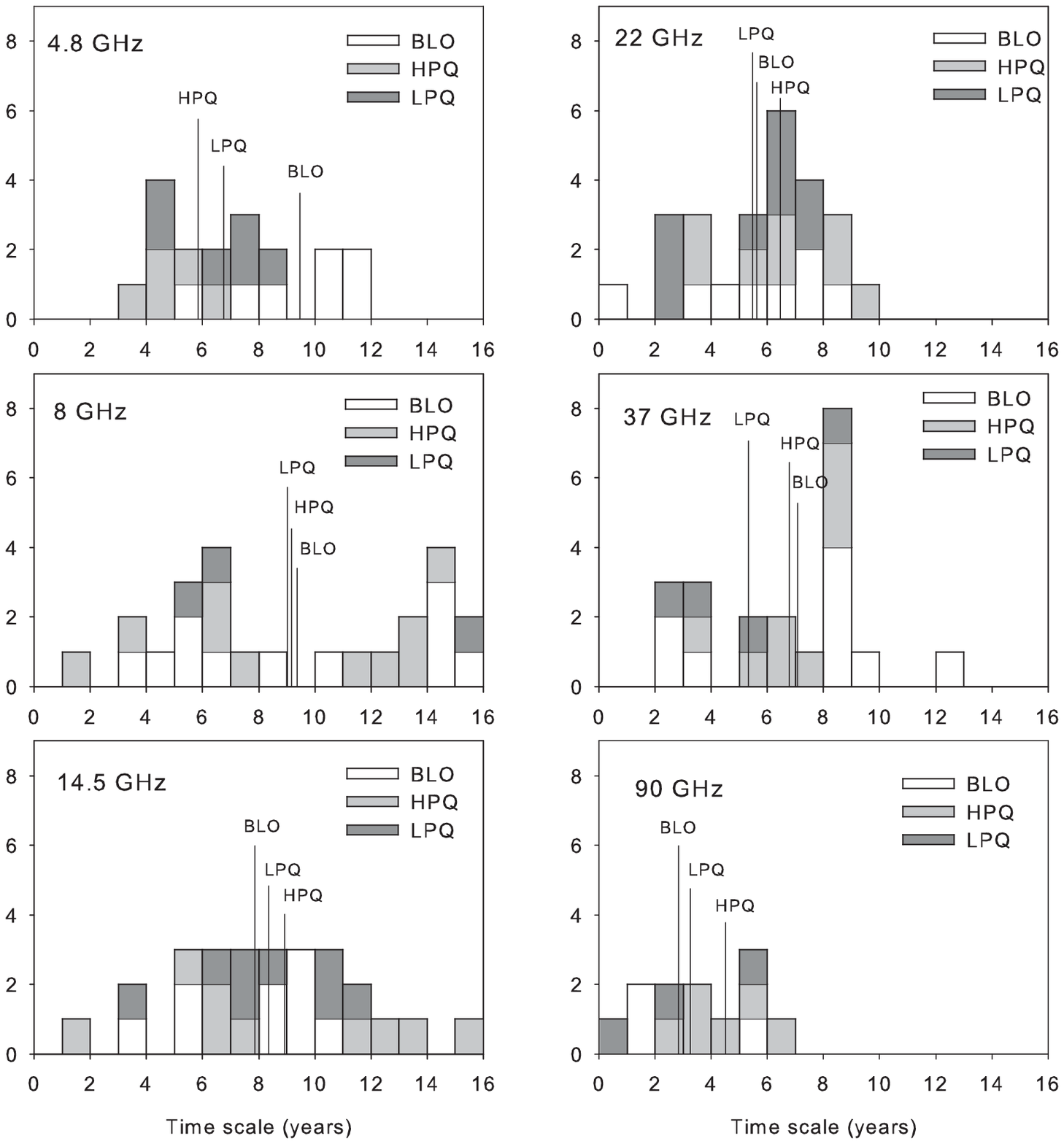}
\caption{Distributions of most significant LS--periodogram time scales at all frequency bands. Averages of each class are shown by vertical lines. The time scales are in the observer's frame.}
\label{histo_period}
\end{figure*}

We also calculated the Lomb-Scargle periodogram for our source sample.
Altogether we obtained 411 periodograms. On 
average we have data at 5 different frequency 
bands for each source. From these 
periodograms we found 140 time scales. In other cases there were no 
significant spikes in the periodogram or the spikes were at time scales 
over half of the total length of the time series. We did not take 
these into account because we are interested in time scales that have 
occurred at least twice during the observing period.
In Fig. \ref{histo_period} we have plotted histograms of the distribution of 
time scales at all the frequency bands and for all the source classes. 

We have calculated the average time scales at all 
the frequency bands. These are 
shown in Table \ref{table:periodogram_averages}. 
We present the results for the whole sample and also separately
for BLOs, HPQs and LPQs. For each source class
also the number of sources used to calculate the average are given. 
Each frequency band is listed separately, and the averages of the most 
significant time scales are shown as well as the redshift corrected time 
scales.

\begin{table*}
\caption{Averages of the periodogram time scales. For each frequency band the most significant observational and redshift corrected time scale averages are
given.}
\label{table:periodogram_averages}
\centering
\begin{tabular}{rrrrrrrrrr}
\hline
\hline
Freq 	  & 	 type 	  & 	 ALL 	  & 	 number 	  & 	 BLO 	  & 	 number 	  & 	 HPQ 	  & 	 number 	  & 	 LPQ 	  & 	 number 	  \\
 	  & 	  	  & 	 [years] 	  & 	 of sources 	  & [years]	  	  & 	 of sources 	  & [years]	  	  & 	 of sources 	  &[years] 	  	  & 	 of sources 	  \\
\hline
\hline 
 4.8 	  & 	 most signif. 	  & 	 7.6 	  & 	 20 	  & 	 9.4 	  & 	 7 	  & 	 5.9 	  & 	 6 	  & 	 6.7 	  & 	 6 	  \\
  	  & 	 redshift corr. 	  & 	 4.6 	  & 	 20 	  & 	 6.2 	  & 	 7 	  & 	 2.9 	  & 	 6 	  & 	 3.6 	  & 	 6 	  \\
\hline 
 8 	  & 	 most signif. 	  & 	 9.2 	  & 	 25 	  & 	 9.4 	  & 	 11 	  & 	 9.1 	  & 	 10 	  & 	 9.0 	  & 	 3 	  \\
  	  & 	 redshift corr. 	  & 	 5.6 	  & 	 25 	  & 	 6.4 	  & 	 11 	  & 	 5.3 	  & 	 10 	  & 	 3.8 	  & 	 3 	  \\
\hline 
 14.5 	  & 	 most signif. 	  & 	 8.5 	  & 	 29 	  & 	 7.9 	  & 	 9 	  & 	 8.9 	  & 	 9 	  & 	 8.3 	  & 	 8 	  \\
  	  & 	 redshift corr. 	  & 	 5.4 	  & 	 29 	  & 	 5.6 	  & 	 9 	  & 	 5.5 	  & 	 9 	  & 	 4.0 	  & 	 8 	  \\
\hline 
 22 	  & 	 most signif. 	  & 	 5.8 	  & 	 28 	  & 	 5.6 	  & 	 8 	  & 	 6.4 	  & 	 8 	  & 	 5.4 	  & 	 9 	  \\
  	  & 	 redshift corr. 	  & 	 3.9 	  & 	 28 	  & 	 4.7 	  & 	 8 	  & 	 3.6 	  & 	 8 	  & 	 3.0 	  & 	 9 	  \\
\hline 
 37 	  & 	 most signif. 	  & 	 6.3 	  & 	 24 	  & 	 7.1 	  & 	 9 	  & 	 6.8 	  & 	 8 	  & 	 5.2 	  & 	 4 	  \\
  	  & 	 redshift corr. 	  & 	 4.3 	  & 	 24 	  & 	 5.6 	  & 	 9 	  & 	 3.7 	  & 	 8 	  & 	 3.3 	  & 	 4 	  \\
\hline 
\end{tabular}
\end{table*}

\subsection{Differences between frequency bands and classes}
We ran a set of Kruskal-Wallis 
tests to search for 
statistically 
significant differences between the individual frequency bands and classes.
The analysis was done for all three analysis methods separately. 
Significant differences between frequencies were found from the 
results of all analysis methods.

In all cases the lower frequencies from 4.8 to 14.5\,GHz formed one group.
In the same way 22 and 37\,GHz formed their own group. 
In addition there were groups with 
only higher frequency data from 22\,GHz upwards, including 90 and 230\,GHz 
or a group with either 22 or 37\,GHz with 4.8\,GHz. 

We did not find significant differences between the classes in most of the 
frequency bands. In SF analysis only at 4.8\,GHz LPQs differed from the other 
classes significantly. Also in LS--periodogram we could find differences only 
at 4.8\,GHz, but this time it was the HPQs and BLOs which were drawn from 
different populations. In the DCF analysis we did not find any significant 
differences between the source classes.

\subsection{Redshift corrections}
We have made redshift corrections for the time scales of all those sources 
in our sample for which we could find a redshift in the literature. We did not 
have the redshift for only one BL Lac object, and for two other  
BLOs we had only a lower limit estimate. The redshift corrected averages 
for the analyses are shown in Tables \ref{table:SF_averages}, 
\ref{table:DCF_averages} and \ref{table:periodogram_averages}

In the SF analysis we found that at 22 and 37\,GHz, the BLOs and 
the HPQs were drawn from different populations. At other frequency bands 
we did not find any significant differences. 
In the DCF analysis at 4.8\,GHz the HPQs and BLOs differed from each other and 
at 37\,GHz the BLOs differed from both quasar types significantly. In the
LS--periodogram the results were quite similar, only at 4.8\,GHz could we 
find any significant differences with BLOs differing from quasars. The 
results indicated that there may be a difference between the quasars and 
BLOs also at other frequency bands but there were only a few objects in each 
class and 
we would need more objects to draw conclusions about the differences. For 
example at 37\,GHz the probability for HPQs and LPQs being from the same 
population was much higher than the probability of BLOs being from the same 
population with either of them.

\subsection{Correlations}
We have studied the correlation between the redshift corrected time scales from 
DCF analysis and different jet parameters including the Lorentz factor,
the Doppler boosting factor and the viewing angle. We have taken the values 
from \cite{lahteenmaki99b}, where the different parameters have been 
calculated for 81 sources. The Doppler boosting factors 
are calculated from total flux density variations and for each source the flare 
with the highest intrinsic brightness temperature is chosen 
\citep{lahteenmaki99}. This usually means strong and rapid flares. 

The samples have 64 
sources in common and for those sources we 
used the Doppler boosting factors provided.
The Lorentz factors and viewing angles were calculated for 
41 sources in common with our sample. In our calculations of the luminosities 
we have used the 
same cosmology as in \cite{lahteenmaki99b} with 
$\mathrm{H_0} = 100 \mathrm{km s^{-1} Mpc^{-1}}$ and $\mathrm{q_0} = 0.5$.

We have not corrected the time scales for Doppler boosting, because the 
time scales from DCF analysis are time intervals between the flares and 
therefore Doppler boosting should not affect them.

\begin{figure}
\resizebox{\hsize}{!}{\includegraphics{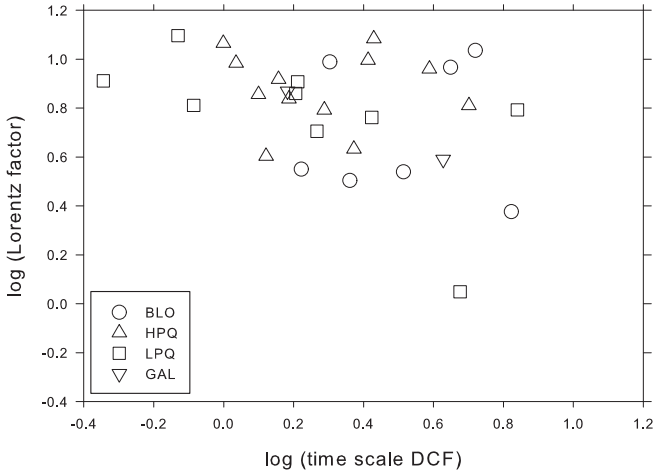}}
\caption{Correlation between the Lorentz factors and the time scales from DCF analysis at 37\,GHz.}
\label{37_lorentz}
\end{figure}

Figure \ref{37_lorentz} shows
Lorentz factor against the time scale from DCF analysis at 37\,GHz. 
We had 30 sources for which both the Lorentz factor and the time scale 
were determined. We calculated the Spearman Rank Correlation and found a 
significant correlation of $r=-0.37$ between the parameters. 
There was, however, one source (2021+614) affecting the correlation 
with a very small Lorentz factor of 
1.12. Ignoring this source made the correlation insignificant.
We have also made a linear fit to the log-log values of 
the data by using ordinary least-squares bisector, suitable for data with both 
X and Y errors unknown \citep{isobe90}. The slope from the fit is -0.87 which 
would indicate that the time scales t are related to 
the Lorentz factor $\Gamma$ as $\mathrm{t}\propto\Gamma^{-1.2}$.

\begin{figure}
\resizebox{\hsize}{!}{\includegraphics{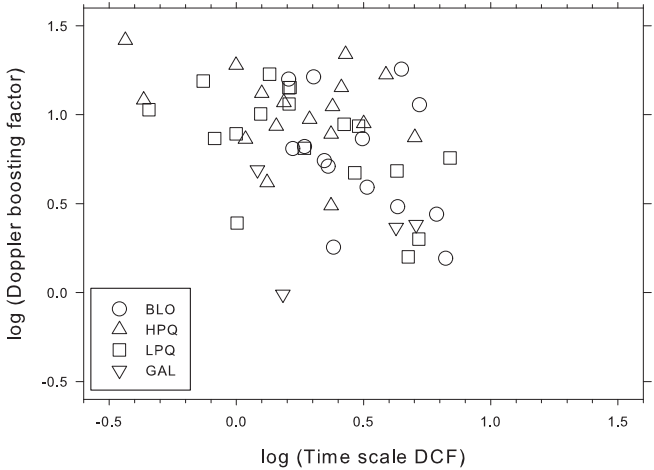}}
\caption{Correlation between the Doppler factors and the time scales from 
DCF analysis at 37\,GHz.}
\label{37_doppler}
\end{figure}

A similar fit can be made for the Doppler boosting factors 
and the time scales (Fig. \ref{37_doppler}) using 53 sources at 37\,GHz. 
The result is quite surprising, a dependence with a slope of -1.1 is 
found implying a relation of $\mathrm{t}\propto\mathrm{D}^{-0.93}$. 
The correlation coefficient is $r= -0.41$, which is significant 
at P$>$99\% level. 

\begin{figure}
\resizebox{\hsize}{!}{\includegraphics{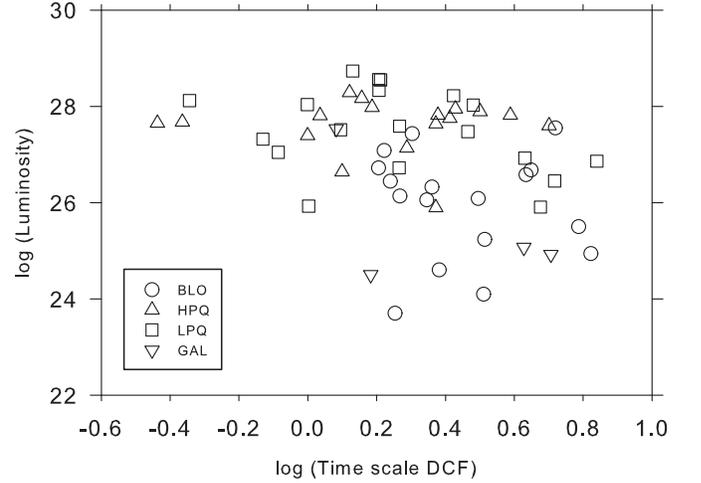}}
\caption{Correlation between the absolute luminosity and the time scale from DCF analysis at 37\,GHz.}
\label{Lum_Tscale}
\end{figure}

The correlation between the absolute luminosity of the source 
and intrinsic time scales at 37 GHz is shown in Fig. \ref{Lum_Tscale}. 
The correlation coefficient is $r=-0.39$, which is significant at 
99\% level. We obtain a slope of -2.8 from the linear fit resulting in a
relation of $\mathrm{t}\propto\mathrm{L}^{-0.35}$ 
between the time scale t and the luminosity L of the source.

\begin{figure}
\resizebox{\hsize}{!}{\includegraphics{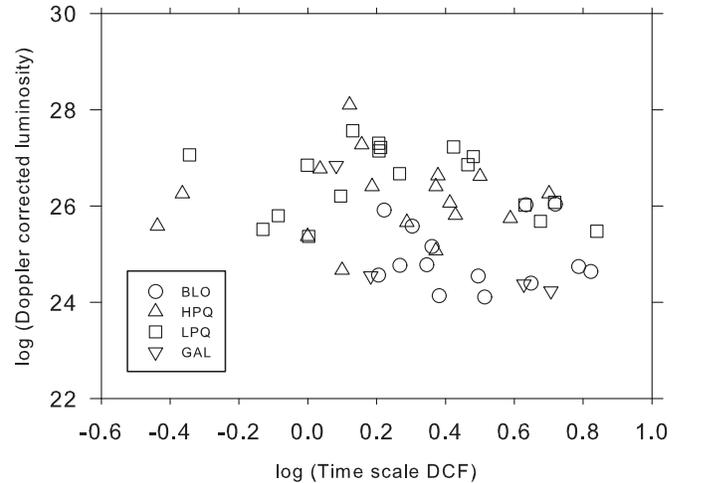}}
\caption{Correlation between the Doppler corrected absolute luminosity and the time scales from DCF analysis at 37\,GHz.}
\label{LumD2_Tscale}
\end{figure}

According to the standard model, Doppler boosting also affects the 
luminosity. Figure \ref{LumD2_Tscale} shows the D$^2$ corrected luminosity 
against the time scale. 
The relation we obtain from a linear fit is 
$\mathrm{t}\propto\mathrm{L}^{-0.46}$. 
The correlation coefficient is $r=-0.33$, which is significant at 
99\% level. 

We do not find any significant correlation between the viewing angle 
and the time scales.

We got quite similar results when we made the fits for 22\,GHz data. The 
slopes were slightly different and also the number of sources for which we 
had both the parameter and the time scale determined was lower. Therefore 
we consider the 37\,GHz results more reliable.

\section{Discussion}\label{sec:discussion}
\subsection{Comparison of the structure function analyses}
We have compared the results from these analyses with 
the results from Paper I.
Table \ref{table:SF_param} lists the slopes and time scales 
for the 40 sources that were common to both studies.
Two of the sources in the sample of Paper I
are not included in this paper, because they had not been 
monitored intensively after 
1993. 
In some cases the slope was complicated, and a representative value 
of the slope is appended with an index c.

\begin{table*}
\caption[]{Parameters of the Structure Functions. 
Slopes
which were complicated or more difficult to determine are marked with letter c.
Those for which we were unable to determine the slope or the time scale are 
marked with -. Time scales are given in years.
}
\label{table:SF_param}
\centering
\begin{tabular}{r|llrr|llrr}
\hline
\hline
Name 	  & 	 22\,GHz 	  & 	 22\,GHz 	  & 	 22\,GHz 	  & 	 22\,GHz 	  & 	 37\,GHz 	  & 	 37\,GHz 	  & 	 37\,GHz 	  & 	 37\,GHz 	  \\
 	  & 	 b 	  & 	 b old 	  & 	 $T_\mathrm{max}$ 	  & 	 $T_\mathrm{max}$ old 	  & 	 b 	  & 	 b old 	  & 	 $T_\mathrm{max}$ 	  & 	 $T_\mathrm{max}$ old 	  \\
\hline
\hline
 \object{0007$+$106} 	  & 	 1.2 	  & 	 1.2 	  & 	 1.08 	  & 	 1.06 	  & 	 0.9 	  & 	 0.9 	  & 	 2.15 	  & 	 1.78 	  \\
 \object{0106$+$013} 	  & 	 0.9 	  & 	 1.4 	  & 	 2.42 	  & 	 $>6.68$ 	  & 	 0.8 	  & 	 1.4 	  & 	 1.36 	  & 	 $>8.41$ 	  \\
 \object{0133$+$476} 	  & 	 $1^c$ 	  & 	 0.7 	  & 	 $>9.62$ 	  & 	 2.11 	  & 	 0.5 	  & 	 0.8 	  & 	 0.76 	  & 	 1.68 	  \\
 \object{0235$+$164} 	  & 	 1.1 	  & 	 1.5 	  & 	 0.96 	  & 	 0.94 	  & 	 0.8 	  & 	 1 	  & 	 2.71 	  & 	 0.94 	  \\
 \object{0248$+$430} 	  & 	 0.4 	  & 	 2.1 	  & 	 $>13.58$ 	  & 	 $>5.62$ 	  & 	 0 	  & 	 2.3 	  & 	 - 	  & 	 $>7.50$ 	  \\
 \object{0316$+$413} 	  & 	 1.8 	  & 	 1.5 	  & 	 $>21.53$ 	  & 	 $>9.44$ 	  & 	 2 	  & 	 1.5 	  & 	 3.83 	  & 	 3.76 	  \\
 \object{0333$+$321} 	  & 	 0.6 	  & 	 0.8 	  & 	 1.92 	  & 	 $>3.76$ 	  & 	 0.7 	  & 	 0.9 	  & 	 1.21 	  & 	 1.06 	  \\
 \object{0355$+$508} 	  & 	 1.2 	  & 	 1.5 	  & 	 7.64 	  & 	 $>10.59$ 	  & 	 $1.6^c$ 	  & 	 1.7 	  & 	 $>12.11$ 	  & 	 $>9.44$ 	  \\
 \object{0420$-$014} 	  & 	 $0.9^c$ 	  & 	 0.9 	  & 	 4.3 	  & 	 1.06 	  & 	 1 	  & 	 0.8 	  & 	 1.36 	  & 	 1.68 	  \\
 \object{0422$+$004} 	  & 	 0.6 	  & 	 1 	  & 	 $>9.62$ 	  & 	 $>4.22$ 	  & 	 0.6 	  & 	 1.5 	  & 	 2.71 	  & 	 $>4.22$ 	  \\
 \object{0430$+$052} 	  & 	 1.1 	  & 	 0.8 	  & 	 0.43 	  & 	 0.6 	  & 	 1 	  & 	 0.7 	  & 	 0.54 	  & 	 1.5 	  \\
 \object{0458$-$020} 	  & 	 1 	  & 	 1.2 	  & 	 $>9.62$ 	  & 	 0.67 	  & 	 $1^c$ 	  & 	 1.3 	  & 	 1.71 	  & 	 1.88 	  \\
 \object{0642$+$449} 	  & 	 0.6 	  & 	 - 	  & 	 $>7.64$ 	  & 	 1 	  & 	 0.4 	  & 	 0.9 	  & 	 $>8.57$ 	  & 	 2.99 	  \\
 \object{0735$+$178} 	  & 	 1.3 	  & 	 1 	  & 	 6.07 	  & 	 4.22 	  & 	 1 	  & 	 1.1 	  & 	 2.15 	  & 	 3.76 	  \\
 \object{0736$+$017} 	  & 	 1.4 	  & 	 0.7 	  & 	 0.22 	  & 	 1.19 	  & 	 1.4 	  & 	 0.9 	  & 	 0.27 	  & 	 0.53 	  \\
 \object{0754$+$100} 	  & 	 1 	  & 	 1 	  & 	 0.54 	  & 	 1.88 	  & 	 0.6 	  & 	 1.3 	  & 	 1.21 	  & 	 1.5 	  \\
 \object{0851$+$202} 	  & 	 $0.9^c$ 	  & 	 1 	  & 	 0.3 	  & 	 0.42 	  & 	 0.7 	  & 	 1.1 	  & 	 0.48 	  & 	 0.11 	  \\
 \object{0923$+$392} 	  & 	 1.3 	  & 	 1.3 	  & 	 $>13.58$ 	  & 	 $>10.59$ 	  & 	 1.4 	  & 	 1.3 	  & 	 5.41 	  & 	 $>5.31$ 	  \\
 \object{1055$+$018} 	  & 	 1.2 	  & 	 1.2 	  & 	 1.36 	  & 	 $>9.44$ 	  & 	 0.8 	  & 	 0.9 	  & 	 1.21 	  & 	 $>10.59$ 	  \\
 \object{1156$+$295} 	  & 	 0.9 	  & 	 1.1 	  & 	 1.21 	  & 	 $>6.68$ 	  & 	 1 	  & 	 2.1 	  & 	 1.36 	  & 	 $>4.73$ 	  \\
 \object{1219$+$285} 	  & 	 0.3 	  & 	 1.1 	  & 	 $>21.53$ 	  & 	 $>10.59$ 	  & 	 0.4 	  & 	 1.8 	  & 	 4.3 	  & 	 $>9.44$ 	  \\
 \object{1226$+$023} 	  & 	 1.3 	  & 	 1.4 	  & 	 1.36 	  & 	 1.5 	  & 	 1.3 	  & 	 1.5 	  & 	 1.21 	  & 	 1.19 	  \\
 \object{1253$-$055} 	  & 	 $1.2^c$ 	  & 	 2.4 	  & 	 0.54 	  & 	 0.67 	  & 	 0.9 	  & 	 1.5 	  & 	 0.76 	  & 	 0.75 	  \\
 \object{1308$+$326} 	  & 	 0.8 	  & 	 1 	  & 	 2.71 	  & 	 3.35 	  & 	 0.8 	  & 	 1 	  & 	 3.04 	  & 	 3.73 	  \\
 \object{1418$+$546} 	  & 	 0.5 	  & 	 - 	  & 	 $>7.64$ 	  & 	 - 	  & 	 0.4 	  & 	 1.1 	  & 	 0.61 	  & 	 0.94 	  \\
 \object{1510$-$089} 	  & 	 1 	  & 	 0.9 	  & 	 0.3 	  & 	 1.33 	  & 	 1 	  & 	 1 	  & 	 0.61 	  & 	 1.19 	  \\
 \object{1538$+$149} 	  & 	 0.4 	  & 	 1.6 	  & 	 $>7.64$ 	  & 	 $>7.50$ 	  & 	 0.4 	  & 	 3.9 	  & 	 $>7.64$ 	  & 	 $>7.50$ 	  \\
 \object{1633$+$382} 	  & 	 0.8 	  & 	 2.3 	  & 	 1.52 	  & 	 2.37 	  & 	 0.9 	  & 	 0.7 	  & 	 1.52 	  & 	 $>8.41$ 	  \\
 \object{1641$+$399} 	  & 	 1.1 	  & 	 1.2 	  & 	 1.36 	  & 	 1.33 	  & 	 1.2 	  & 	 1.4 	  & 	 1.36 	  & 	 1.19 	  \\
 \object{1741$-$038} 	  & 	 1.2 	  & 	 1.3 	  & 	 1.36 	  & 	 $>4.22$ 	  & 	 0.6 	  & 	 1.4 	  & 	 3.83 	  & 	 $>4.22$ 	  \\
 \object{1749$+$096} 	  & 	 1.2 	  & 	 0.7 	  & 	 0.61 	  & 	 $>5.96$ 	  & 	 1.2 	  & 	 0.6 	  & 	 0.34 	  & 	 $>10.59$ 	  \\
 \object{1807$+$698} 	  & 	 0.2 	  & 	 - 	  & 	 - 	  & 	 - 	  & 	 0 	  & 	 - 	  & 	 - 	  & 	 - 	  \\
 \object{2005$+$403} 	  & 	 0.8 	  & 	 0.9 	  & 	 $>13.58$ 	  & 	 $>10.59$ 	  & 	 0.9 	  & 	 0.9 	  & 	 6.07 	  & 	 $>9.44$ 	  \\
 \object{2134$+$004} 	  & 	 0.4 	  & 	 1.2 	  & 	 1.92 	  & 	 $>5.96$ 	  & 	 0.4 	  & 	 1.3 	  & 	 3.41 	  & 	 3.76 	  \\
 \object{2145$+$067} 	  & 	 1.3 	  & 	 0.9 	  & 	 1.36 	  & 	 $>4.22$ 	  & 	 1 	  & 	 0.8 	  & 	 1.36 	  & 	 $>4.22$ 	  \\
 \object{2200$+$420} 	  & 	 $0.9^c$ 	  & 	 1.1 	  & 	 2.42 	  & 	 0.67 	  & 	 0.9 	  & 	 1.1 	  & 	 0.48 	  & 	 0.38 	  \\
 \object{2201$+$315} 	  & 	 1.3 	  & 	 1.5 	  & 	 1.52 	  & 	 $>2.99$ 	  & 	 1.2 	  & 	 1.5 	  & 	 3.41 	  & 	 $>3.76$ 	  \\
 \object{2223$-$052} 	  & 	 1 	  & 	 1.3 	  & 	 2.71 	  & 	 1.5 	  & 	 1 	  & 	 1.5 	  & 	 3.04 	  & 	 1.06 	  \\
 \object{2230$+$114} 	  & 	 1.2 	  & 	 0.7 	  & 	 1.71 	  & 	 0.53 	  & 	 0.8 	  & 	 1.2 	  & 	 6.07 	  & 	 0.38 	  \\
 \object{2251$+$158} 	  & 	 0.9 	  & 	 1 	  & 	 2.15 	  & 	 2.37 	  & 	 1 	  & 	 1.5 	  & 	 0.54 	  & 	 0.53 	  \\
\hline 
\end{tabular}
\end{table*}

We had a total of 75 source/frequency combinations where the 
slope and the time scale, or at least its lower limit, 
were determined from both analyses.
Many sources had changed their behaviour during the years so that 
the results in the two studies differed greatly from each other. 
This indicates that even 10 years of monitoring is 
not enough to reveal the true nature of all the sources. There were also 
sources whose behaviour had stayed the same during the 25 years.
In 20 cases the time scale had changed less than one year between 
the two studies and the slope had changed less than 0.3.
In a total of 35 cases the time scale had changed less than a year.
In 44, slightly more than half of the cases, the slope had changed 
$\leq 0.3$.
In 21 cases the time scale had changed more than 
3 years, but four of these were sources for which only a lower 
limit estimate could be determined in both analyses. In those cases the 
longer observing period caused the time scale to change more even though the 
behaviour of the source had not changed.

Paper I had 32 cases which gave only a lower limit for the variability time
scale. In our analysis the number was reduced to fourteen. 
This is not unexpected as the total observing 
time in the present study is twice as long as in the previous analysis.

The time scales in our analysis are in general shorter than in Paper I. 
At 22\,GHz the median value had changed from 3.0 
to 1.9 years and at 37\,GHz from 3.0 to 1.4 years. This can be partially 
explained with the smaller number of lower limit estimates in our analysis
than in Paper I. Also, there are many sources that exhibit much more dramatic
variability behaviour in the data set obtained after Paper I. 
Naturally there are also sources for which exactly the opposite
is true, but because the SF determines the shortest time
scale from the complete data sets, we now obtained altogether
a larger number of short timescales than in Paper I.
The time scales at 37\,GHz 
are also shorter than those at 22\,GHz, possibly indicating that some faster
flux density variations are not clearly distinguishable at lower 
frequencies, including 22\,GHz.

The average values for the slopes in our analysis were 
close to unity,
typical for shot noise. At 22\,GHz the average slope is 0.96 and at 
37\,GHz 0.86. These are slightly smaller than the values in the previous
paper, where the average at 22\,GHz was 1.2 and at 37\,GHz 1.3. The difference 
between the slopes in our analysis and Paper I was significant at the 99\% level.
Paper I predicted 
that the number of very steep
structure functions should be smaller with longer monitoring because
usually such sources have few prominent outbursts which dominate the
structure function. This effect should be smaller after longer
monitoring with more flares seen in the flux density curves. This can be
clearly seen from our analysis where there are only two sources with a
slope over 1.5 whereas in Paper I there were 13 sources with a slope $\geq$1.5.

We did not find statistically significant differences in the time scales 
between the source classes LPQ, HPQ and BLO. The results in paper I 
suggested a statistically significant difference in the time scales of LPQs 
and HPQs, but we cannot confirm this. 

\subsection{Comparison of the time series analysis methods}
We also compared 
the different methods used for
studying the time scales. The structure function gives a different time
scale compared to the other two methods. 
The structure function focuses
on the characteristic time scales of the flux density variations,
for example, the rise and decay times of the flares.
Similar conclusion was made by \cite{tanihata01} where they used the 
SF method for studying the X-ray flares of AGNs. 
The DCF and the periodogram focus more on the
periodicity and quasi-periodicity in the flux density curves and are more
affected by 
the time cycle of 
the large outbursts. We could also use the SF to study the time scales 
between the flares by examining the first minimum in the second plateau of 
the SF instead of the point where the plateau starts. The minima are 
usually easier to see if the SF is plotted in linear scale. Here we have 
compared the SF time scales with Paper I and therefore used the logarithmic 
scale and the different time scale.

When examining the flux density curves, 
the time scales obtained from the DCF or the LS--periodogram could usually be 
identified with the time intervals between some 
large flares.
This is also why the time scales from the DCF and
the LS--periodogram are longer than the ones obtained from the 
structure function analysis. There is usually some variation in
the flux density levels also in the short time 
scales but the big outbursts occur
quite rarely as can be seen from the averages of the DCF and
the LS--periodogram analysis.

The time scales are shorter at 4.8\,GHz than at 
8\,GHz with every method we used. 
This can be 
due to lower flux density levels. According to the general shock model 
\citep[for example][]{valtaoja92},
the flares last longer at the lower frequencies 
and the growing and decaying shocks overlap, 
forming a smoother curve. The flare peaks are not as extreme as at
higher frequencies, and the statistical methods catch smaller
peaks.   
Another explanation could be that the 8\,GHz band has been monitored 
for longer periods than the 4.8\,GHz band. The time scales are therefore 
longer, because we see more large events in the flux curves and they occur 
more rarely.

We could determine both a DCF and a LS--periodogram time scale in 136 cases.
In one third of the cases the methods gave the same result within 0.5 years. 
In half of the cases the difference 
in the time scales was less than a year.
\begin{figure}
\resizebox{\hsize}{!}{\includegraphics{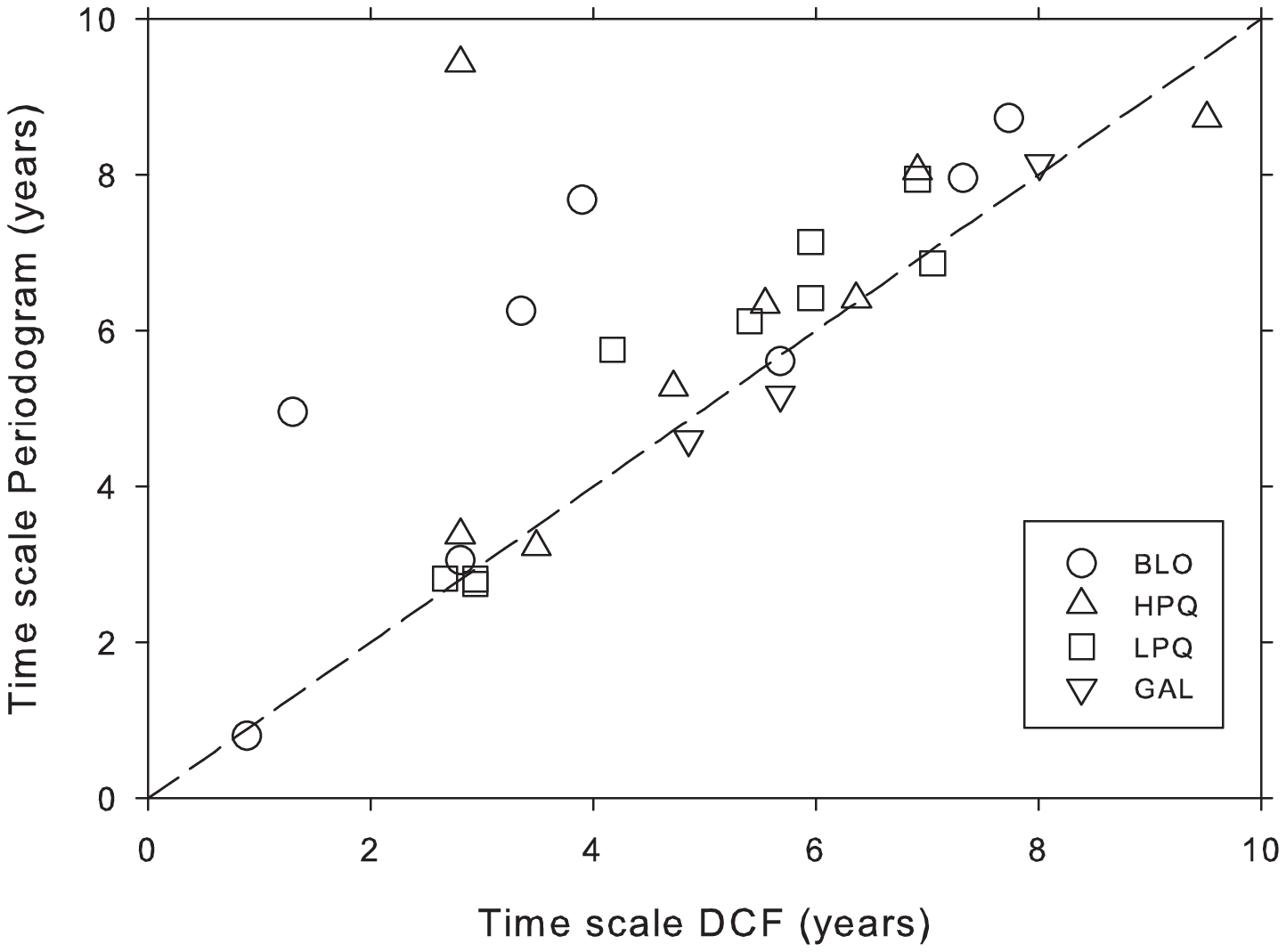}}
\caption{Time scales from the LS analysis plotted against the time scales from the DCF analysis at 22\,GHz. Time scales are equal on the dashed line.}
\label{LS_vs_DCF}
\end{figure}

In Fig. \ref{LS_vs_DCF} we have plotted the most significant time scale
from the LS analysis against the time scale from the DCF analysis at 22\,GHz. 
There is clearly a linear 
one-to-one correspondence 
between the time scales. 
There are also a few time scales above the line, and in these cases 
the DCF gives a shorter time scale.
In all of these cases the DCF shows multiple correlations and time scales.
In every case the DCF 
shows another time
scale within a year of the periodogram time scale, but has it not been 
the first correlation, which was our criterion for the DCF time scale, 
it is not taken into account. Therefore we can conclude that the two methods 
give very similar results.

\begin{figure}
\resizebox{\hsize}{!}{\includegraphics{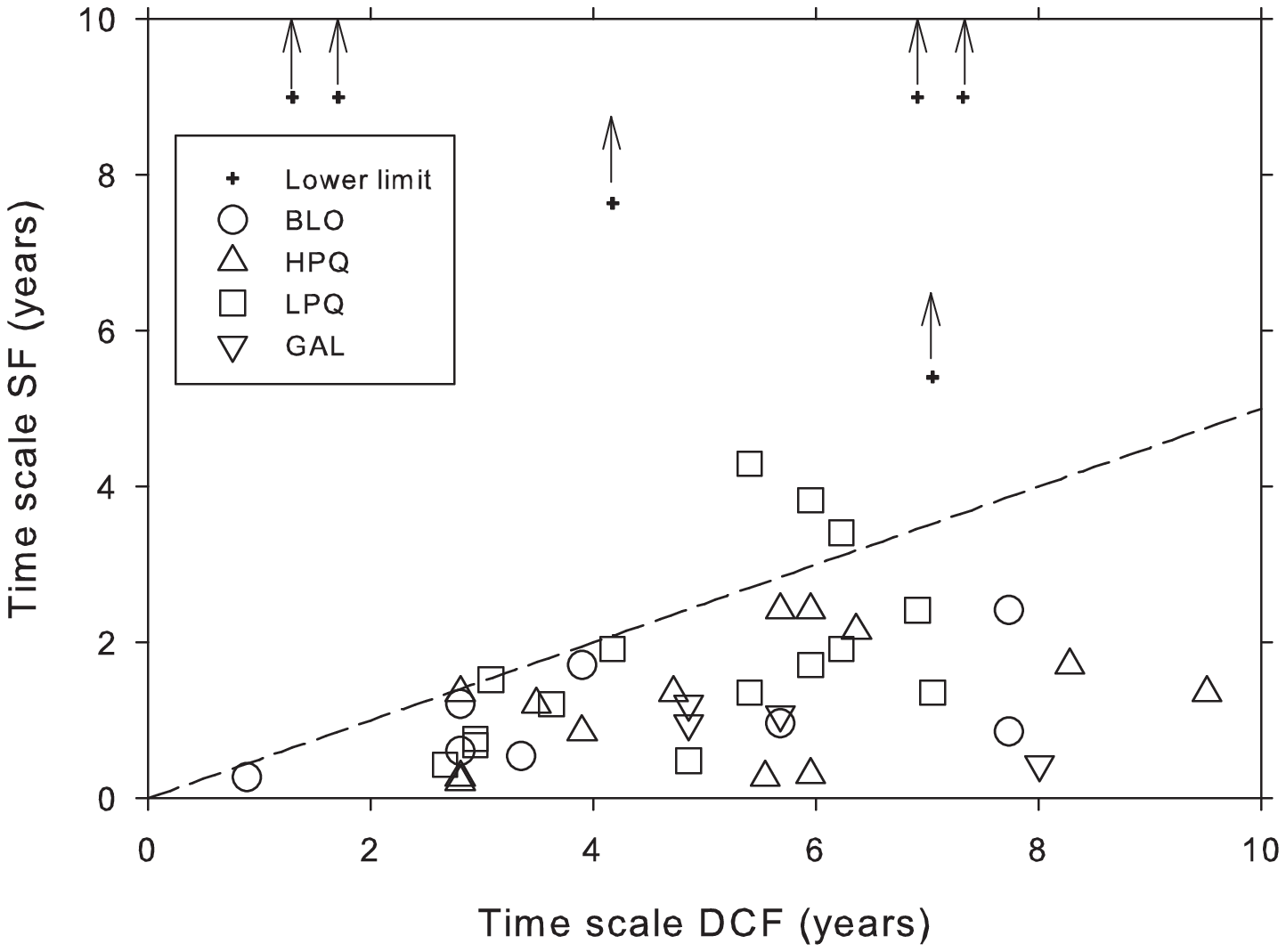}}
\caption{ Time scales of the SF analysis plotted against the time scales of the DCF at 22\,GHz. Time scales for which the structure function could only determine a lower limit are marked with crosses. The lower limits of time scales in the upper part are from left to right 12.1, 12.1, 10.8 and 21.5 years. Dashed line shows the SF time scale equals half of the DCF time scale line.
}
\label{LS_vs_SF}
\end{figure}

In Fig. \ref{LS_vs_SF} 
we have plotted time scales from the SF analysis against the time scales 
from the DCF analysis.
The lower limits of time scales from 
SF analysis are plotted using crosses. When ignoring these time 
scales we can see that the SF time scales are 
one half of the DCF time scales or shorter.
This is expected since the SF should see the rise and 
decay times of flares whereas the DCF sees the time interval between them or the total length of the flare. 

The differences can be explained with a following example: 
We have a pure noiseless sinusoid with a period of $P$. 
The frequency $(1/P)$ found by the periodogram gives us a direct measure of
the period. This is the time scale at which all the power of the variability is
focused. In a DCF one is searching for a maximum correlation beyond the trivial
timelag of $\tau=0$. Such a maximum is found at $\tau=P$, and all integer 
multiples of $P$. This is a time scale that is in full agreement with the one 
obtained from the periodogram. In a structure function the philosophy for the 
search of the time scale is somewhat different. Here we are searching for a 
time scale at which a {\it maximum} difference between the original and the 
time lagged light curves are found. For a periodic function this is reached 
when the timelag is $\tau=P/2$. Note that at a time scale $\tau=P$, in our 
simple example a deep minimum is achieved in the SF, but in reality this is 
filled to some extent by noise and other non-period features.

The fact that the SF is interpreted in a way to measure the maximum of the
variance whereas DCF and periodogram measure the minimum in variance 
explains the factor 2 difference in the estimated time scales. Our data does 
not have strict periods, but still the different nature of the analyses 
reflect the property that periodogram and DCF give more characteristic 
measures of the times between the outbursts whereas the SF
measure the rise/decay time scales, and thus a factor of 2 or larger is 
expected in the ratio of the estimated time scales.

\begin{figure*}
\centering
\includegraphics[width=17cm, height=15cm]{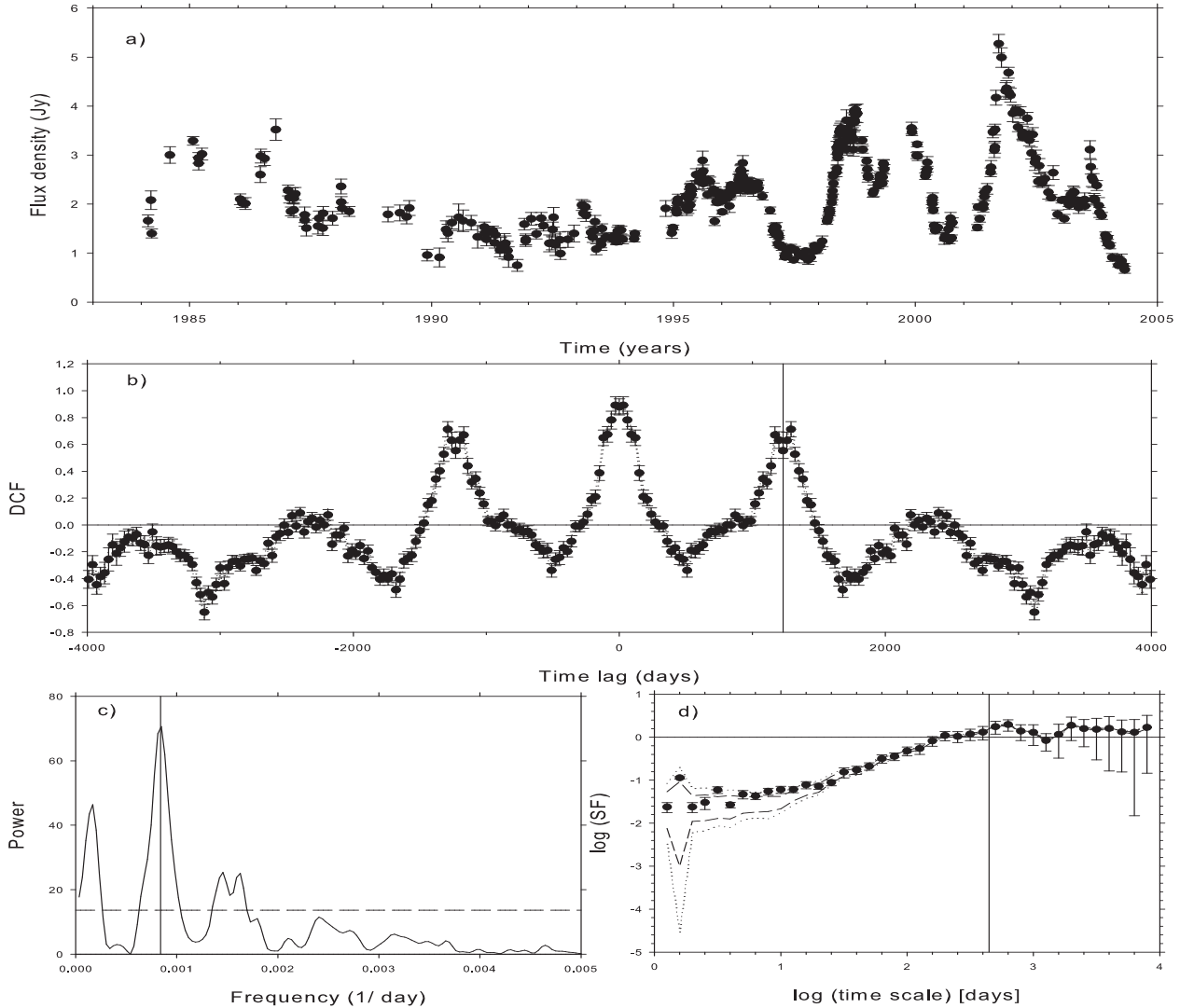}
\caption{Analyses of the HPQ source 1156+295 at 22\,GHz. a) Flux Density curve. b) The discrete correlation function. The 99.5\% significance level is shown with dotted line. c) The Lomb-Scargle periodogram. Dashed line shows the false-alarm probability. d) The structure function. Dashed and dotted lines show the 97.5\% and 99.5\% significance levels. Time scales obtained with each method are marked by vertical lines. The most significant spike of the periodogram is at time scale of 3.29 years, which is 0.2 years shorter than the first correlation in the DCF at 3.49 years. The SF gives a time scale of 1.21 years.
}
\label{1156+295_example}
\end{figure*}

As an example, Fig. \ref{1156+295_example} shows the results of all analysis
methods for a HPQ source \object{1156+295} (\object{4C29.45}). This 
source is a good example of both the DCF and the LS--periodogram 
giving the same time scale within 0.2 years of each other. 
The LS--periodogram 
is plotted in Fig. \ref{1156+295_example}c. We can easily define the 
most significant spike at a 3.29 years time scale. The DCF is plotted in 
Fig. \ref{1156+295_example}b, and there is only one clear correlation at 
a 3.49 years time scale. Both of the plots are easy to interpret and the 
results seem reasonable when examining the flux density curve in Fig. 
\ref{1156+295_example}a. There are larger outbursts with approximately 
3.5 years between them.

The SF, which is plotted in Fig. \ref{1156+295_example}d, gives a shorter 
time scale of 1.21 years. 
By examining the flux density curve, we can see that the rise and decay times of the 
flares are around one year. If we compare the results with Paper I, 
we notice that in Paper I the time scale obtained for this source is over 
6.68 years. The difference between the old and new results can also be
explained by examining the flux density curve. Before 1993 the monitoring was not as 
dense as it has been later and the smaller changes have not been observed
because of the sparser sampling. Also, 1156+295 appears to have changed 
its behaviour in 1995. 

\begin{figure*}
\centering
\includegraphics[width=17cm, height=15cm]{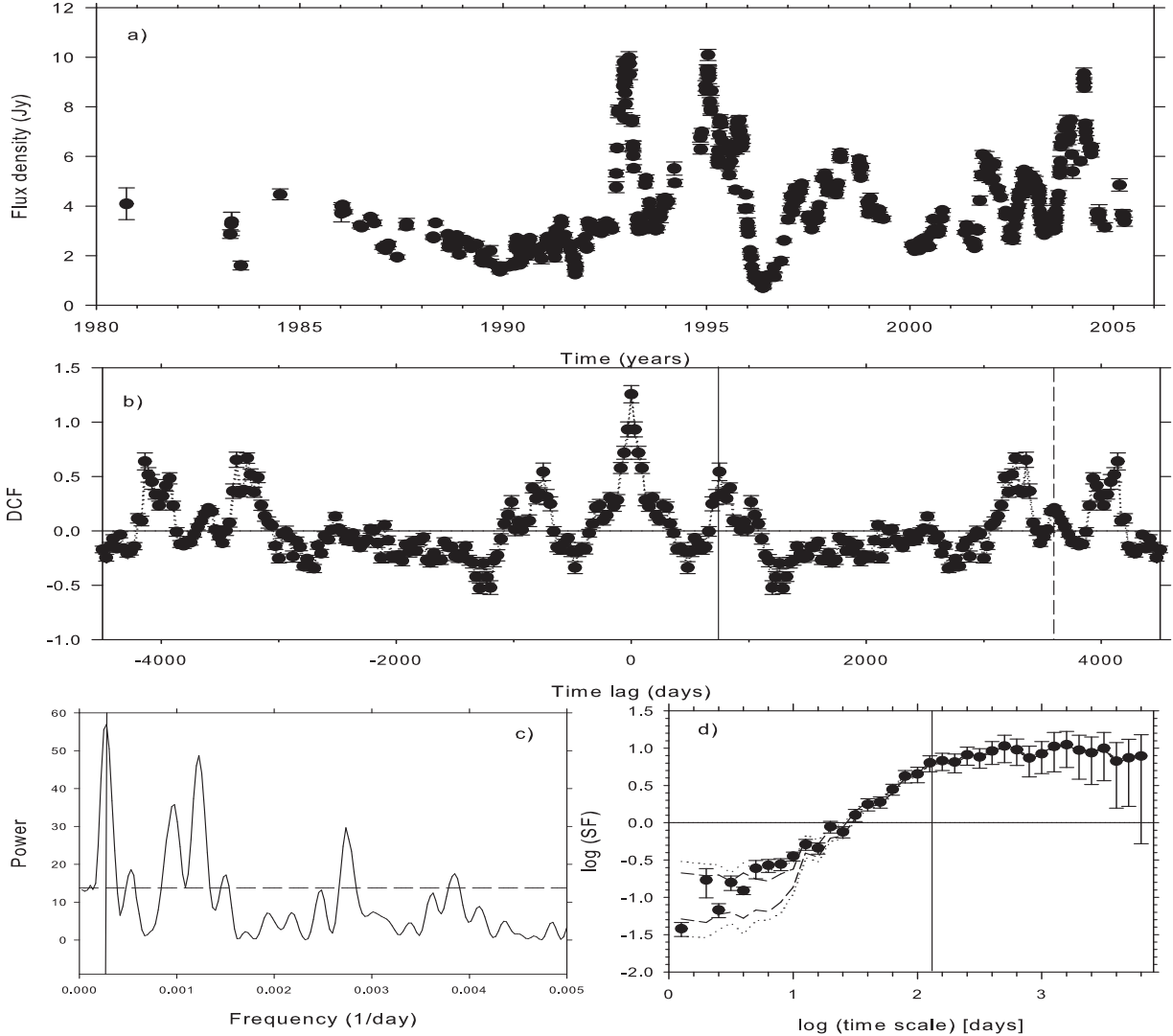}
\caption{Analyses of the BLO source 1749+096 at 37\,GHz. a) Flux Density curve. b) The discrete correlation function. The 99.5\% significance level is shown with dotted line. c) The Lomb-Scargle periodogram. Dashed line shows the false-alarm probability. d) The structure function. Dashed and dotted lines show the 97.5\% and 99.5\% significance levels. Time scales obtained with each method are marked by vertical lines. The most significant spike of the periodogram is at time scale of 9.81 years, which is the same as the DCF correlation marked with vertical dashed line at 9.79 years time scale. The most significant DCF time scale is at 2.12 years. The SF gives a time scale of 0.34 years.
}
\label{1749+096_example}
\end{figure*}

Another example, Fig. \ref{1749+096_example}, showing multiple 
DCF correlations is a BLO source \object{1749+096}. The most significant time 
scale of the DCF is at 2.12 years, but there are multiple correlations and 
one at the time scale of 9.79 years, corresponding to the time scale 
obtained with the LS--periodogram at 9.81 years. Again the SF gives a much 
shorter time scale of 0.34 years, which is due to the short rise times of 
flares at 1993 and 1995. The source has clearly changed its behaviour from 
the time of Paper I, where the time scale obtained is more than 10.59 years.  

The results were not always this well in 
agreement
with each other, and 
occasionally, it was rather difficult to find the true time scales. Usually 
this happens when there are gaps in the data or the flux density curve had
a strong linear baseline.
Mostly, however, the results were quite 
easily obtained and they matched 
well the behaviour seen visually in the flux density curves. 

\subsection{Intrinsic time scales and the correlations}
The DCF and the periodogram give an average time of about three years 
between the flares. This is the time interval between the shocks in the jet. 
Although there were indications that BLOs and 
quasars may differ from each other, the differences are very small 
considering how different the jets in BLOs and quasars are thought to be
\citep[e.g.][]{aller99}.

It is also interesting to notice how little the time scale varies compared 
to the change in the jet parameters and luminosity. Even though the observed
luminosities of 
these jets varied five orders of magnitude, the variability behaviour is very 
similar in all sources. Also, when we take the Doppler effect into account, 
the luminosities vary four orders of magnitude. 
We could expect the luminosity
and the time scales to have a stronger dependence as one would expect 
the luminosity and 
the formation of the shocks to depend on the mass and the accretion rate
of the central black hole.
This was not seen in our analysis and it strengthens 
the idea that the shocks are formed by jet instabilities rather than the 
central engine itself. The same conclusion could be drawn from the result 
that the time scales are 
inversely proportional to luminosity meaning that greater luminosity sources 
have shorter time scales. One might think that in larger sources things
happen more slowly and the time scales are longer as predicted by the 
sequence from microquasars to Low Luminosity
AGNs to quasars, which gives a variability time scale that is proportional
to the mass of the central black hole. Our analysis 
shows the opposite. Also \cite{aller06} found no correlation between the 
radio variability and the black hole mass for the MOJAVE sample.

The correlations also show that the change in time scale compared to the 
change in Lorentz factor is quite small. This is actually quite surprising 
if we think that the jet instabilities, which cause the shocks, are 
strongly related to the speed of the flow. Therefore in the future we need to 
compare the time scales with other parameters affecting the shock formation.
For this we need VLBI information of a large sample of sources.

\section{Conclusions}\label{sec:conclusions}
We have studied the variability time scales of a large sample of sources with 
different statistical methods. In our analyses we used data at frequencies 
from 4.8 to 230\,GHz. One aim was to study how 13 more years of 
data affect the time scales by comparing our results with the analysis of 
Paper I. Many of the sources had changed their behaviour during this time and 
the time scales we obtained by using the structure function differed greatly 
from the ones in Paper I. This shows that even 10 years of monitoring is not 
enough to reveal the true variability behaviour of all the sources. 

In many sources we could see continuous variability, for example small flares,  
but larger flares occur quite 
rarely. For some sources even 25 years is not long enough a time to 
reveal any characteristic time scales of variability at all.

The different methods should be used for different purposes. 
The LS--periodogram and the DCF give the time scale between flares and 
the structure function a characteristic time scale for the source, for example
the rise or decay time scales of flares. When studying the correlations 
between the time scales and different jet parameters, we considered the 
DCF time scale to be the most reliable in giving the time scale between the 
flares. The LS--periodogram produces easily spurious spikes and therefore 
needs to be used with caution. In this paper also a method for studying the 
significance levels of the DCF was developed which in turn made the analysis 
more reliable.

We did not find any significant differences between the various source classes
when either directly observed or redshift corrected  time scales were considered. 
There 
was an indication that the BLOs may differ from quasars when intrinsic 
time scales are considered, but the differences are modest.

The range in time scales compared to the range of luminosities was 
very small indicating that the shock formation is not strongly related to 
the mass and the accretion rate of the central black hole but instead may be 
related to the jet instabilities.

\begin{acknowledgements}
We acknowledge the support of the Academy of Finland for
the Mets\"ahovi and SEST observing projects.
UMRAO is supported in part by a series of grants from the NSF and by funds 
from the University of Michigan Department of Astronomy.
\end{acknowledgements}

\bibliographystyle{aa}
\bibliography{7529}

\onltab{1}{
\longtab{1}{
\begin{longtable}{lll|ll|ll|ll}
\caption[]{\label{table:sourcelist} List of sources and frequencies. For each frequency the length of the time series and the number of observations are given.}\\
\hline
\hline
Name 	  &  	  & 	 Class 	  & 	 22\,GHz 	  & 	  	  & 	 37\,GHz 	  & 	  	  & 	 90\,GHz 	  & 	  	  \\
 	  & 	  	  & 	  	  & 	 years  	  & 	 N 	  & 	years  	  & 	 N	  & 	 years 	  & 	 N	  \\
\hline 
\endfirsthead 
\caption{continued.}\\
\hline\hline
Name 	  & 	   & 	 Class 	  & 	 22\,GHz 	  & 	  	  & 	 37\,GHz 	  & 	  	  & 	 90\,GHz 	  & 	  	  \\
 	  & 	  	  & 	  	  & 	 years 	  & 	 N	  & 	 years  	  & 	 N 	  & 	 years 	  & 	 N 	  \\
\hline
\endhead
\hline
\endfoot
 \object{0007+106} 	  & 	 \object{III ZW 2} 	  & 	 GAL 	  & 	 19.409 	  & 	 309 	  & 	 19.230 	  & 	 253 	  & 	  	  & 	  	  \\
 \object{0016+731} 	  & 	  	  & 	 LPQ 	  & 	 11.875 	  & 	 72 	  & 	 10.806 	  & 	 62 	  & 	  	  & 	  	  \\
 \object{0106+013} 	  & 	 \object{OC 012} 	  & 	 HPQ 	  & 	 22.275 	  & 	 197 	  & 	 24.155 	  & 	 160 	  & 	  	  & 	  	  \\
 \object{0109+224} 	  & 	 \object{S2 0109+22} 	  & 	 BLO 	  & 	 19.332 	  & 	 181 	  & 	 20.657 	  & 	 151 	  & 	  	  & 	  	  \\
 \object{0133+476} 	  & 	 \object{DA 55} 	  & 	 HPQ 	  & 	 22.433 	  & 	 335 	  & 	 24.165 	  & 	 259 	  & 	 14.924 	  & 	 73 	  \\
 \object{0149+218} 	  & 	  	  & 	 LPQ 	  & 	 15.891 	  & 	 147 	  & 	 16.687 	  & 	 97 	  & 	  	  & 	  	  \\
 \object{0202+149} 	  & 	 \object{4C 15.05} 	  & 	 HPQ 	  & 	 19.931 	  & 	 216 	  & 	 20.657 	  & 	 187 	  & 	  	  & 	  	  \\
 \object{0212+735} 	  & 	  	  & 	 HPQ 	  & 	 15.965 	  & 	 103 	  & 	 16.739 	  & 	 55 	  & 	  	  & 	  	  \\
 \object{0224+671} 	  & 	  	  & 	 LPQ 	  & 	 13.372 	  & 	 106 	  & 	 14.866 	  & 	 79 	  & 	  	  & 	  	  \\
 \object{0234+285} 	  & 	 \object{4C 28.07} 	  & 	 HPQ 	  & 	 16.172 	  & 	 145 	  & 	 15.079 	  & 	 105 	  & 	 10.260 	  & 	 80 	  \\
 \object{0235+164} 	  & 	  	  & 	 BLO 	  & 	 22.436 	  & 	 399 	  & 	 24.112 	  & 	 540 	  & 	 12.312 	  & 	 160 	  \\
 \object{0248+430} 	  & 	  	  & 	 LPQ 	  & 	 17.590 	  & 	 92 	  & 	 20.726 	  & 	 107 	  & 	  	  & 	  	  \\
 \object{0316+413} 	  & 	 \object{3C 84} 	  & 	 GAL 	  & 	 22.447 	  & 	 911 	  & 	 25.460 	  & 	 1360 	  & 	 14.950 	  & 	 265 	  \\
 \object{0333+321} 	  & 	 \object{NRAO 140} 	  & 	 LPQ 	  & 	 17.277 	  & 	 178 	  & 	 25.370 	  & 	 154 	  & 	  	  & 	  	  \\
 \object{0336-019} 	  & 	 \object{CTA 026} 	  & 	 HPQ 	  & 	 15.362 	  & 	 108 	  & 	 16.717 	  & 	 82 	  & 	 12.281 	  & 	 67 	  \\
 \object{0355+508} 	  & 	 \object{NRAO 150} 	  & 	 LPQ 	  & 	 22.428 	  & 	 327 	  & 	 25.408 	  & 	 268 	  & 	 14.950 	  & 	 102 	  \\
 \object{0415+379} 	  & 	 \object{3C 111} 	  & 	 GAL 	  & 	 11.504 	  & 	 147 	  & 	 12.458 	  & 	 88 	  & 	  	  & 	  	  \\
 \object{0420-014} 	  & 	 \object{OA 129} 	  & 	 HPQ 	  & 	 22.319 	  & 	 378 	  & 	 21.112 	  & 	 417 	  & 	 14.925 	  & 	 198 	  \\
 \object{0422+004} 	  & 	 \object{OF 038} 	  & 	 BLO 	  & 	 19.253 	  & 	 144 	  & 	 19.202 	  & 	 130 	  & 	  	  & 	  	  \\
 \object{0430+052} 	  & 	 \object{3C 120} 	  & 	 GAL 	  & 	 22.419 	  & 	 417 	  & 	 24.123 	  & 	 461 	  & 	 14.950 	  & 	 108 	  \\
 \object{0446+112} 	  & 	 \object{PKS 0446+112} 	  & 	 GAL 	  & 	 15.967 	  & 	 119 	  & 	 16.717 	  & 	 70 	  & 	  	  & 	  	  \\
 \object{0458-020} 	  & 	 \object{PKS 0458-020} 	  & 	 HPQ 	  & 	 16.163 	  & 	 106 	  & 	 17.077 	  & 	 86 	  & 	  	  & 	  	  \\
 \object{0528+134} 	  & 	 \object{PKS 0528+134} 	  & 	 LPQ 	  & 	 15.962 	  & 	 548 	  & 	 16.919 	  & 	 371 	  & 	 13.469 	  & 	 130 	  \\
 \object{0552+398} 	  & 	 \object{DA 193} 	  & 	 LPQ 	  & 	 14.045 	  & 	 217 	  & 	 14.956 	  & 	 171 	  & 	 13.406 	  & 	 75 	  \\
 \object{0642+449} 	  & 	 \object{OH 471} 	  & 	 LPQ 	  & 	 23.747 	  & 	 270 	  & 	 24.101 	  & 	 219 	  & 	  	  & 	  	  \\
 \object{0716+714} 	  & 	  	  & 	 BLO 	  & 	 15.905 	  & 	 203 	  & 	 16.742 	  & 	 492 	  & 	 8.820 	  & 	 84 	  \\
 \object{0735+178} 	  & 	 \object{PKS 0735+17} 	  & 	 BLO 	  & 	 23.747 	  & 	 309 	  & 	 24.038 	  & 	 295 	  & 	 14.942 	  & 	 99 	  \\
 \object{0736+017} 	  & 	  	  & 	 HPQ 	  & 	 21.220 	  & 	 208 	  & 	 21.902 	  & 	 157 	  & 	 11.466 	  & 	 79 	  \\
 \object{0754+100} 	  & 	 \object{OI 090.4} 	  & 	 BLO 	  & 	 20.329 	  & 	 208 	  & 	 25.367 	  & 	 170 	  & 	  	  & 	  	  \\
 \object{0804+499} 	  & 	  	  & 	 HPQ 	  & 	 15.970 	  & 	 257 	  & 	 16.744 	  & 	 160 	  & 	  	  & 	  	  \\
 \object{0814+425} 	  & 	  	  & 	 BLO 	  & 	 16.117 	  & 	 199 	  & 	 16.885 	  & 	 122 	  & 	  	  & 	  	  \\
 \object{0827+243} 	  & 	 \object{OJ 248} 	  & 	 LPQ 	  & 	 10.728 	  & 	 133 	  & 	 11.361 	  & 	 62 	  & 	  	  & 	  	  \\
 \object{0836+710} 	  & 	 \object{4C 71.07} 	  & 	 LPQ 	  & 	 16.060 	  & 	 224 	  & 	 16.866 	  & 	 226 	  & 	  	  & 	  	  \\
 \object{0851+202} 	  & 	 \object{OJ 287} 	  & 	 BLO 	  & 	 23.266 	  & 	 895 	  & 	 24.824 	  & 	 912 	  & 	 14.955 	  & 	 272 	  \\
 \object{0906+430} 	  & 	 \object{3C 216} 	  & 	 HPQ 	  & 	 20.318 	  & 	 74 	  & 	 21.926 	  & 	 53 	  & 	  	  & 	  	  \\
 \object{0923+392} 	  & 	 \object{4C 39.25} 	  & 	 LPQ 	  & 	 23.865 	  & 	 773 	  & 	 24.787 	  & 	 542 	  & 	 14.905 	  & 	 103 	  \\
 \object{0945+408} 	  & 	 \object{4C 40.24} 	  & 	 LPQ 	  & 	 15.819 	  & 	 113 	  & 	 16.727 	  & 	 55 	  & 	  	  & 	  	  \\
 \object{0953+254} 	  & 	  	  & 	 LPQ 	  & 	 15.964 	  & 	 156 	  & 	 16.706 	  & 	 133 	  & 	  	  & 	  	  \\
 \object{0954+556} 	  & 	 \object{S4 0954+556} 	  & 	 HPQ 	  & 	 15.816 	  & 	 108 	  & 	 16.704 	  & 	 55 	  & 	  	  & 	  	  \\
 \object{0954+658} 	  & 	 \object{S4 0954+65} 	  & 	 BLO 	  & 	 16.137 	  & 	 107 	  & 	 21.897 	  & 	 78 	  & 	  	  & 	  	  \\
 \object{1055+018} 	  & 	 \object{OL 093} 	  & 	 HPQ 	  & 	 22.430 	  & 	 255 	  & 	 24.076 	  & 	 229 	  & 	 12.744 	  & 	 80 	  \\
 \object{1101+384} 	  & 	 \object{MARK 421} 	  & 	 BLO 	  & 	 15.007 	  & 	 360 	  & 	 19.130 	  & 	 275 	  & 	  	  & 	  	  \\
 \object{1147+245} 	  & 	 \object{B2 1147+24} 	  & 	 BLO 	  & 	 15.817 	  & 	 63 	  & 	 15.792 	  & 	 37 	  & 	  	  & 	  	  \\
 \object{1156+295} 	  & 	 \object{4C 29.45} 	  & 	 HPQ 	  & 	 20.185 	  & 	 404 	  & 	 21.920 	  & 	 326 	  & 	 12.520 	  & 	 73 	  \\
 \object{1219+285} 	  & 	 \object{ON 231} 	  & 	 BLO 	  & 	 23.885 	  & 	 281 	  & 	 24.111 	  & 	 212 	  & 	  	  & 	  	  \\
 \object{1222+216} 	  & 	 \object{PKS 1222+216} 	  & 	 LPQ 	  & 	 10.548 	  & 	 243 	  & 	 11.320 	  & 	 116 	  & 	  	  & 	  	  \\
 \object{1226+023} 	  & 	 \object{3C 273} 	  & 	 LPQ 	  & 	 24.030 	  & 	 939 	  & 	 25.303 	  & 	 1039 	  & 	 14.956 	  & 	 352 	  \\
 \object{1253-055} 	  & 	 \object{3C 279} 	  & 	 HPQ 	  & 	 24.006 	  & 	 762 	  & 	 25.303 	  & 	 789 	  & 	 14.925 	  & 	 234 	  \\
 \object{1308+326} 	  & 	 \object{AU CV n} 	  & 	 BLO 	  & 	 22.283 	  & 	 378 	  & 	 24.005 	  & 	 315 	  & 	 10.043 	  & 	 85 	  \\
 \object{1413+135} 	  & 	  	  & 	 BLO 	  & 	 15.365 	  & 	 243 	  & 	 16.134 	  & 	 185 	  & 	 10.116 	  & 	 71 	  \\
 \object{1418+546} 	  & 	 \object{OQ 530} 	  & 	 BLO 	  & 	 21.051 	  & 	 207 	  & 	 21.958 	  & 	 182 	  & 	  	  & 	  	  \\
 \object{1502+106} 	  & 	 \object{OR 103} 	  & 	 HPQ 	  & 	 16.109 	  & 	 156 	  & 	 23.473 	  & 	 136 	  & 	  	  & 	  	  \\
 \object{1510-089} 	  & 	 \object{PKS 1510-089} 	  & 	 HPQ 	  & 	 19.943 	  & 	 245 	  & 	 21.926 	  & 	 263 	  & 	 14.055 	  & 	 111 	  \\
 \object{1538+149} 	  & 	 \object{4C 14.60} 	  & 	 BLO 	  & 	 20.327 	  & 	 236 	  & 	 21.932 	  & 	 181 	  & 	  	  & 	  	  \\
 \object{1606+106} 	  & 	 \object{4C 10.45} 	  & 	 LPQ 	  & 	 11.277 	  & 	 150 	  & 	 12.101 	  & 	 108 	  & 	  	  & 	  	  \\
 \object{1611+343} 	  & 	 \object{DA 406} 	  & 	 LPQ 	  & 	 15.964 	  & 	 229 	  & 	 16.832 	  & 	 198 	  & 	  	  & 	  	  \\
 \object{1633+382} 	  & 	 \object{4C 38.41} 	  & 	 LPQ 	  & 	 22.431 	  & 	 457 	  & 	 24.082 	  & 	 466 	  & 	  	  & 	  	  \\
 \object{1637+574} 	  & 	 \object{OS 562} 	  & 	 LPQ 	  & 	 20.175 	  & 	 150 	  & 	 21.939 	  & 	 156 	  & 	  	  & 	  	  \\
 \object{1641+399} 	  & 	 \object{3C 345} 	  & 	 HPQ 	  & 	 23.998 	  & 	 806 	  & 	 24.800 	  & 	 783 	  & 	 14.933 	  & 	 182 	  \\
 \object{1652+398} 	  & 	 \object{MARK 501} 	  & 	 BLO 	  & 	 16.111 	  & 	 324 	  & 	 16.891 	  & 	 218 	  & 	  	  & 	  	  \\
 \object{1725+044} 	  & 	 \object{PKS 1725+044} 	  & 	 LPQ 	  & 	 12.701 	  & 	 90 	  & 	 13.623 	  & 	 70 	  & 	  	  & 	  	  \\
 \object{1739+522} 	  & 	 \object{S4 1739+52} 	  & 	 HPQ 	  & 	 15.841 	  & 	 149 	  & 	 16.866 	  & 	 121 	  & 	  	  & 	  	  \\
 \object{1741-038} 	  & 	 \object{PKS 1741-038} 	  & 	 HPQ 	  & 	 16.027 	  & 	 272 	  & 	 16.987 	  & 	 295 	  & 	 13.351 	  & 	 111 	  \\
 \object{1749+096} 	  & 	 \object{PKS 1749+096} 	  & 	 BLO 	  & 	 19.858 	  & 	 583 	  & 	 24.532 	  & 	 465 	  & 	 14.099 	  & 	 154 	  \\
 \object{1803+784} 	  & 	 \object{S5 1803+784} 	  & 	 BLO 	  & 	 15.911 	  & 	 120 	  & 	 16.748 	  & 	 103 	  & 	 9.096 	  & 	 106 	  \\
 \object{1807+698} 	  & 	 \object{3C 371.0} 	  & 	 BLO 	  & 	 20.173 	  & 	 143 	  & 	 21.947 	  & 	 137 	  & 	  	  & 	  	  \\
 \object{1823+568} 	  & 	 \object{4C 56.27} 	  & 	 BLO 	  & 	 15.561 	  & 	 61 	  & 	 16.699 	  & 	 35 	  & 	 9.036 	  & 	 63 	  \\
 \object{1928+738} 	  & 	 \object{4C 73.18} 	  & 	 LPQ 	  & 	 16.055 	  & 	 145 	  & 	 16.849 	  & 	 101 	  & 	  	  & 	  	  \\
 \object{2005+403} 	  & 	  	  & 	 LPQ 	  & 	 22.286 	  & 	 318 	  & 	 24.032 	  & 	 302 	  & 	  	  & 	  	  \\
 \object{2007+776} 	  & 	 \object{S5 2007+77} 	  & 	 BLO 	  & 	 12.381 	  & 	 92 	  & 	 16.832 	  & 	 84 	  & 	  	  & 	  	  \\
 \object{2021+614} 	  & 	 \object{OW 637} 	  & 	 LPQ 	  & 	 16.802 	  & 	 107 	  & 	 21.932 	  & 	 115 	  & 	  	  & 	  	  \\
 \object{2134+004} 	  & 	 \object{OX 057} 	  & 	 LPQ 	  & 	 22.272 	  & 	 225 	  & 	 25.374 	  & 	 232 	  & 	  	  & 	  	  \\
 \object{2136+141} 	  & 	  	  & 	 LPQ 	  & 	 15.506 	  & 	 69 	  & 	 18.205 	  & 	 64 	  & 	  	  & 	  	  \\
 \object{2145+067} 	  & 	  	  & 	 LPQ 	  & 	 18.370 	  & 	 551 	  & 	 19.161 	  & 	 496 	  & 	 14.929 	  & 	 134 	  \\
 \object{2200+420} 	  & 	 \object{BL Lac} 	  & 	 BLO 	  & 	 24.009 	  & 	 965 	  & 	 25.443 	  & 	 996 	  & 	 14.962 	  & 	 145 	  \\
 \object{2201+315} 	  & 	 \object{4C 31.63} 	  & 	 LPQ 	  & 	 19.251 	  & 	 280 	  & 	 22.541 	  & 	 251 	  & 	  	  & 	  	  \\
 \object{2223-052} 	  & 	 \object{3C 446} 	  & 	 BLO 	  & 	 19.349 	  & 	 237 	  & 	 19.216 	  & 	 240 	  & 	 14.940 	  & 	 145 	  \\
 \object{2230+114} 	  & 	 \object{CTA 102} 	  & 	 HPQ 	  & 	 19.248 	  & 	 295 	  & 	 19.208 	  & 	 293 	  & 	 14.496 	  & 	 106 	  \\
 \object{2234+282} 	  & 	  	  & 	 HPQ 	  & 	 15.967 	  & 	 71 	  & 	 15.828 	  & 	 32 	  & 	  	  & 	  	  \\
 \object{2251+158} 	  & 	 \object{3C 454.3} 	  & 	 HPQ 	  & 	 24.006 	  & 	 760 	  & 	 24.480 	  & 	 722 	  & 	 14.950 	  & 	 244 	  \\
 
\end{longtable}
}
}
\end{document}